\documentclass[12pt]{article}
\usepackage{amsmath,amssymb,amsthm,amscd}
\usepackage{latexsym}
\usepackage{graphicx}
\usepackage[all]{xy}
\usepackage{fancybox}
\usepackage{color}
\usepackage{colordvi}
\usepackage{axodraw}
\input xy
\xyoption{all}
\usepackage{eucal}

  \makeatletter
\newfont{\footsc}{cmcsc10 at 8truept}
\newfont{\footbf}{cmbx10 at 8truept}
\newfont{\footrm}{cmr10 at 10truept}
\renewcommand{\ps@plain}{%
\renewcommand{\@oddfoot}{\footsc Hopf algebra approach to Feynman diagram calculations,
{\footbf December 6, 2005}\hfil\footrm\thepage}} \makeatother
\begin{document}

%Statements
\newtheorem{theorem}{Theorem}[section]
\newtheorem{proposition}[theorem]{Proposition}
\newtheorem{definition}[theorem]{Definition}
\newtheorem{lemma}[theorem]{Lemma}
\newtheorem{corollary}[theorem]{Corollary}
\newtheorem{prop-def}{Proposition-Definition}[section]
\newtheorem{claim}{Claim}[section]
\newtheorem{remark}[theorem]{Remark}
\newtheorem{example}[theorem]{Example}
\newtheorem{propprop}{Proposed Proposition}[section]
\newtheorem{conjecture}{Conjecture}
\newenvironment{exam}{\begin{example}\rm}{\end{example}}
\newenvironment{rmk}{\begin{remark}\rm}{\end{remark}}

\def\One{\mathbb{I}}

\def\v{\;\raisebox{-1mm}{\epsfysize=5mm\epsfbox{v.eps}}\;}
\def\f{\;\raisebox{-1mm}{\epsfysize=3mm\epsfbox{f.eps}}\;}
\def\g{\;\raisebox{-1mm}{\epsfysize=4mm\epsfbox{g.eps}}\;}
\def\gg{\;\raisebox{-2mm}{\epsfysize=5mm\epsfbox{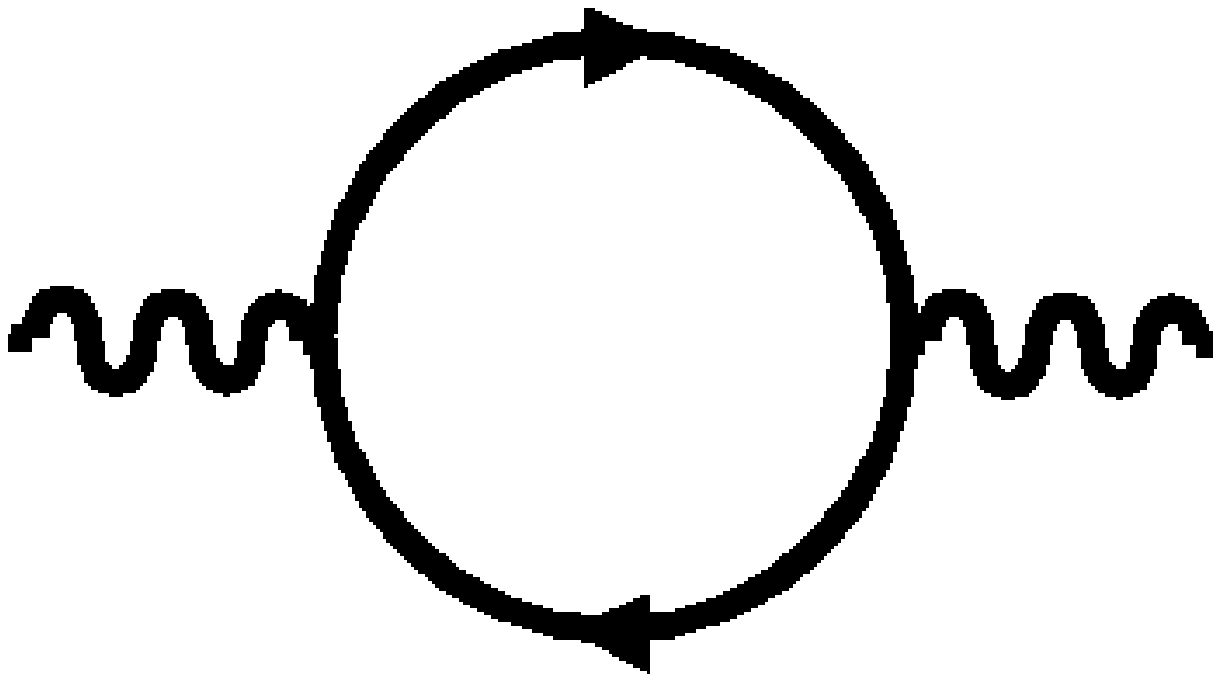}}\;}
\def\ggf{\;\raisebox{-2mm}{\epsfysize=6mm\epsfbox{ggf.eps}}\;}
\def\ggv{\;\raisebox{-3mm}{\epsfysize=8mm\epsfbox{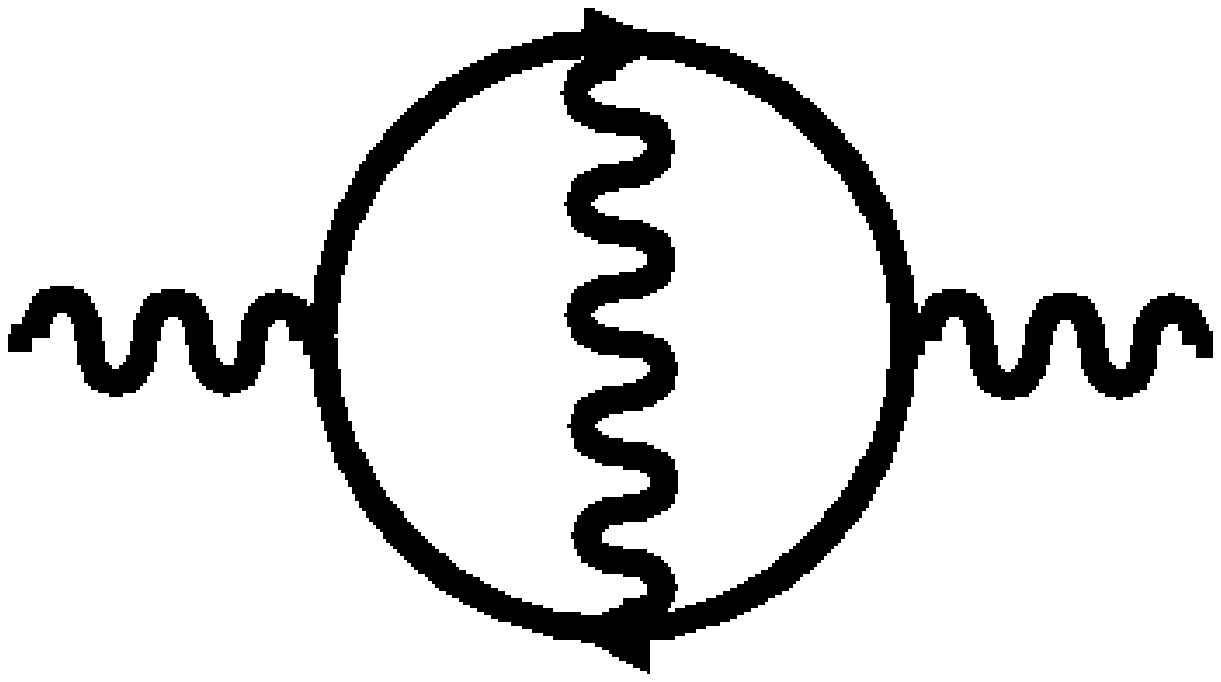}}\;}
\def\epem{\;\raisebox{-1mm}{\epsfysize=3mm\epsfbox{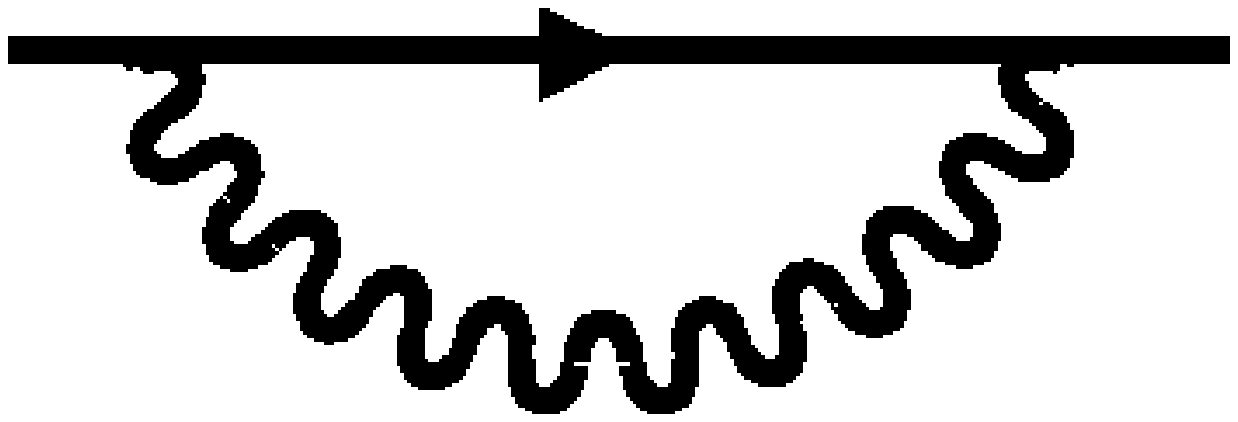}}\;}
\def\epemg{\;\raisebox{-1mm}{\epsfysize=5mm\epsfbox{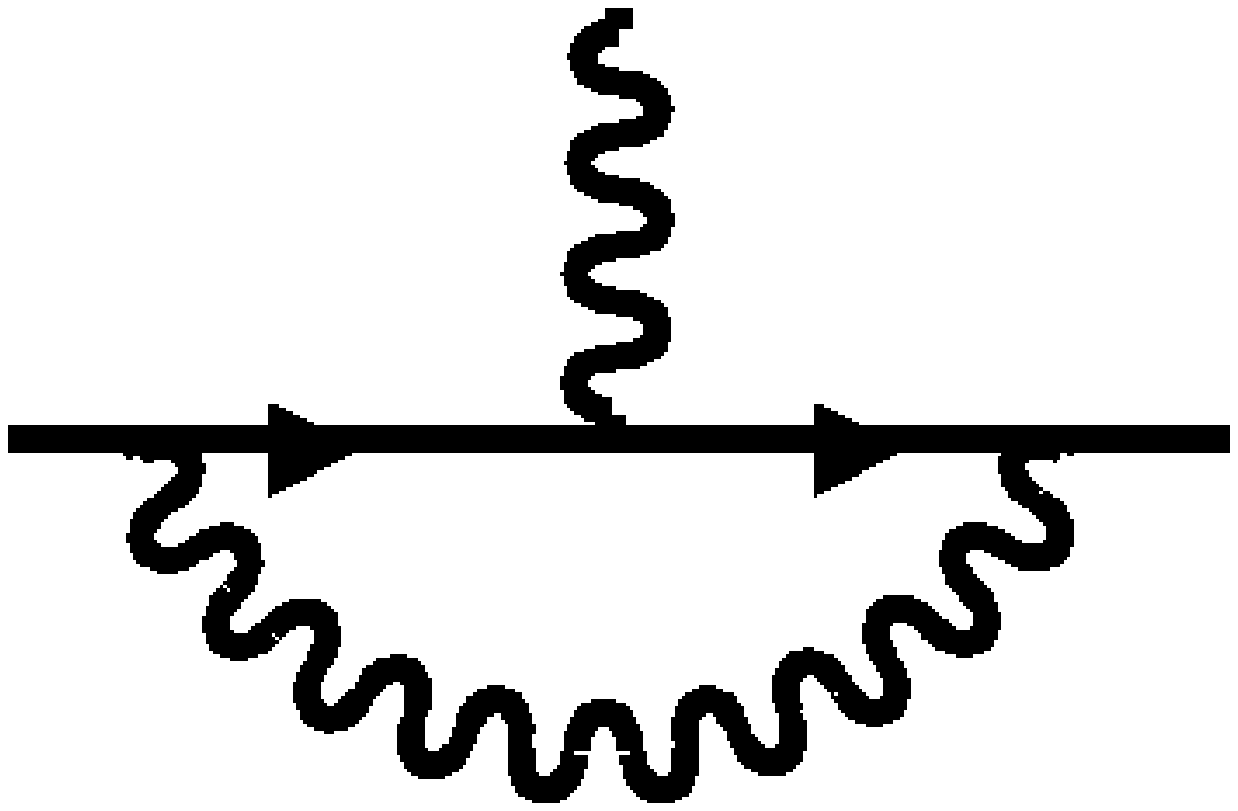}}\;}
\def\epemgg{\;\raisebox{-4mm}{\epsfysize=8mm\epsfbox{epemgg.eps}}\;}
\def\epemgf{\;\raisebox{-3mm}{\epsfysize=8mm\epsfbox{epemgf.eps}}\;}
\def\epemfg{\;\raisebox{-1mm}{\epsfysize=6mm\epsfbox{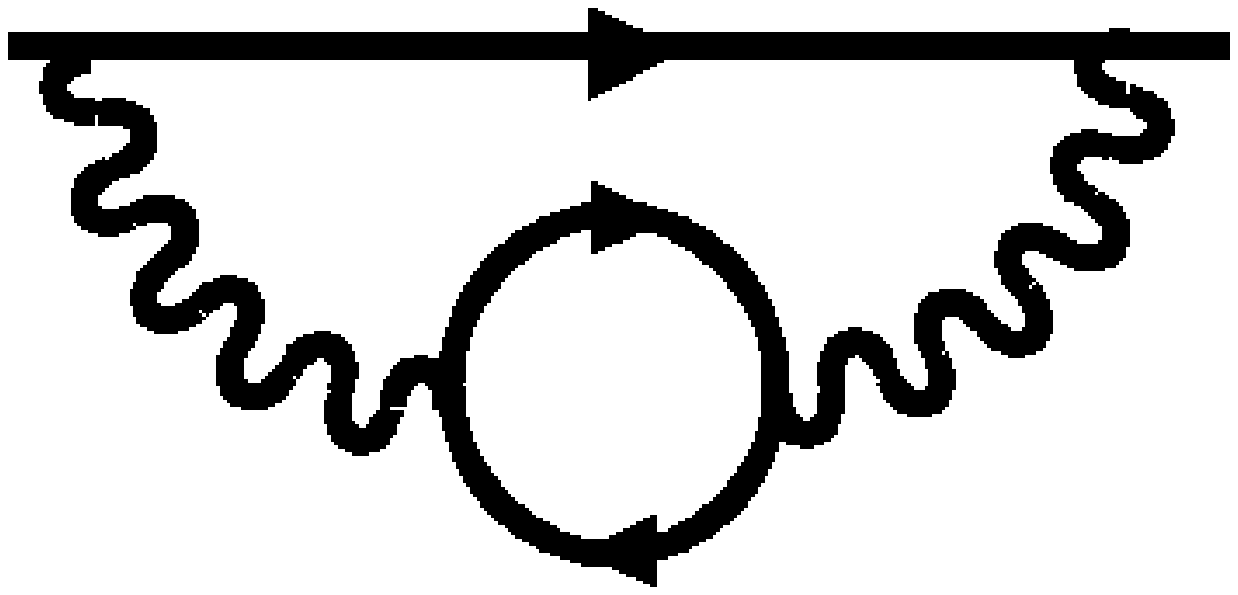}}\;}
\def\epemv{\;\raisebox{-3mm}{\epsfysize=6mm\epsfbox{epemv.eps}}\;}

%%%%%%%%%%%%%%%%%%%%%%%%%%%%%%%%%%%%%%%%%%%%%%%%%%%%%%%%%%%%%%%%%%%QED-graphs%%

\def\FERMprop{{\scalebox{0.4}{ %%%%%%%%%%%%%%%%%%%%%%%%%%%%%%%%%\FERMprop
\begin{picture}(90,0)(30,-45)
\SetWidth{1.5} \SetColor{Black} \ArrowLine(30,-45)(120,-45)
\end{picture}}}}

\def\BOSONprop{{\scalebox{0.4}{ %%%%%%%%%%%%%%%%%%%%%%%%%%%%%%%%%\BOSONprop
\begin{picture}(90,24)(30,-33)
\SetWidth{1.5} \SetColor{Black} \Photon(30,-21)(120,-21){6}{5}
\end{picture}}}}

\def\QEDvertex{{\scalebox{0.4}{ %%%%%%%%%%%%%%%%%%%%%%%%%%%%%%%%%\QEDvertex
\begin{picture}(105,90)(15,-30)
\SetWidth{1.5} \SetColor{Black} \ArrowLine(120,60)(75,15)
\ArrowLine(75,15)(120,-30) \SetWidth{0.5} \Vertex(75,15){5.66}
\SetWidth{1.5} \Photon(75,15)(15,15){6}{5}
\end{picture}}}}

\def\aQEDshrink{{\scalebox{0.4}{ %%%%%%%%%%%%%%%%%%%%%%%%%%%%%%%%%\aQEDshrink
\begin{picture}(166,210)(75,-75)
\SetWidth{1.5} \SetColor{Black} \Photon(75,30)(135,30){5.5}{3}
\Photon(165,60)(165,0){6}{3.5} \ArrowArc(210,30)(31.02,178,538)
\Photon(210,0)(210,-45){6}{2.5} \Photon(210,105)(210,60){6}{2.5}
\ArrowLine(135,30)(165,0) \ArrowLine(165,60)(135,30)
\ArrowLine(240,135)(210,105) \ArrowLine(210,-45)(240,-75)
\ArrowLine(135,30)(240,-75) \ArrowLine(240,135)(135,30)
\ArrowArc(210,30)(30,0,360)
\end{picture}}}}

\def\bQEDshrink{{\scalebox{0.32}{ %%%%%%%%%%%%%%%%%%%%%%%%%%%%%%%%%\bQEDshrink
\begin{picture}(270,270)(45,-30)
\SetWidth{2.0} \SetColor{Black} \Photon(45,105)(105,105){6}{3}
\ArrowLine(270,210)(105,105) \ArrowLine(105,105)(270,0)
\Photon(225,180)(180,105){6}{4} \Photon(165,90)(150,75){6}{1}
\Photon(247,15)(251,196){6}{9} \ArrowLine(105,105)(150,75)
\ArrowLine(150,135)(105,105) \ArrowLine(315,240)(225,180)
\ArrowLine(225,30)(315,-30) \Photon(225,30)(150,135){6}{6.5}
\end{picture}}}}

\def\IQEDse{{\scalebox{0.35}{ %%%%%%%%%%%%%%%%%%%%%%%%%%%%%%%%%\IQEDse
\begin{picture}(246,90)(57,-45)
\SetWidth{2.0} \SetColor{Black} \Photon(57,-1)(132,-1){6}{4}
\Photon(228,0)(303,0){6}{4} \ArrowArc(180,0)(45,270,630)
\ArrowArc(180,0)(45,90,450)
\end{picture}}}}

\def\IIQEDse{{\scalebox{0.35}{ %%%%%%%%%%%%%%%%%%%%%%%%%%%%%%%%%\IIQEDse
\begin{picture}(249,108)(56,-35)
\SetWidth{1.5} \SetColor{Black} \ArrowArc(180,18)(48.84,137,497)
\ArrowArc(180,18)(48.84,47,407) \ArrowArc(180,18)(49.5,225,585)
\ArrowArc(180,18)(48.8,314,674) \Photon(180,66)(180,-31){6}{5}
\SetWidth{2.0} \Photon(230,16)(305,16){6}{4}
\Photon(56,16)(131,16){6}{4}
\end{picture}}}}

\def\IIIQEDse{{\scalebox{0.35}{ %%%%%%%%%%%%%%%%%%%%%%%%%%%%%%%%%\IIIQEDse
\begin{picture}(270,65)(60,-9)
\SetWidth{2.0} \SetColor{Black}
\PhotonArc(195,-10)(60,0,180){6}{9.5} \ArrowLine(255,-10)(330,-10)
\ArrowLine(60,-10)(135,-10) \ArrowLine(135,-10)(255,-10)
\end{picture}}}}

\def\IVQEDse{{\scalebox{0.35}{ %%%%%%%%%%%%%%%%%%%%%%%%%%%%%%%%%\IVQEDse
\begin{picture}(330,130)(60,-9)
\SetWidth{1.5} \SetColor{Black}
\PhotonArc(195,55)(60,0,180){6}{9.5} \SetWidth{2.0}
\ArrowLine(255,55)(330,55) \ArrowLine(60,55)(135,55)
\ArrowLine(135,55)(195,55) \ArrowLine(195,55)(255,55)
\ArrowLine(330,55)(390,55) \PhotonArc(255,55)(60,-180,0){6}{9.5}
\end{picture}}}}

\def\A{\mathcal{A}}
\def\B{\mathcal{B}}
\def\D{\mathcal{D}}
\def\F{\mathcal{F}}
\def\G{\mathcal{G}}
\def\H{\mathcal{H}}
\def\L{\mathcal{L}}
\def\M{\mathcal{M}}
\def\P{\mathcal{P}}
\def\R{\mathcal{R}}
\def\T{\mathcal{T}}
\def\U{\mathcal{U}}
\def\X{\mathcal{X}}

\def\l{\mathfrak{l}}
\def\g{\mathfrak{g}}

\def\CC{\mathbb{C}}
\def\KK{\mathbb{K}}
\def\NN{\mathbb{N}}
\def\QQ{\mathbb{Q}}
\def\RR{\mathbb{R}}
\def\ZZ{\mathbb{Z}}

\newcommand{\id}{\mathrm{id}}
\newcommand{\delete}[1]{}
\newcommand{\End}{\mathrm{End}}
\newcommand{\Hom}{\mathrm{Hom}}

\begin{center}
{\LARGE{\bf Hopf algebra approach to Feynman diagram calculations}}\\[1.5cm]
\end{center}

\begin{center}
         KURUSCH EBRAHIMI-FARD\footnote{fard@th.physik.uni-bonn.de; currently visiting the
         Fields Institute, Toronto, Canada.}\\\smallskip
{\small{
         Universit\"at Bonn -
         Physikalisches Institut\\
         Nussallee 12,
         D-53115 Bonn, Germany}}\\[0.6cm]

         DIRK KREIMER\footnote{kreimer@ihes.fr and dkreimer@bu.edu, Center for Math.Phys.,
         Boston University.}\\\smallskip
{\small{
         C.N.R.S.-I.H.\'E.S.\\
         Le Bois-Marie, 35, Route de Chartres\\
         F-91440 Bures-sur-Yvette, France}}\\[2cm]

December 6, 2005\\[1.5cm]

    {\texttt{J. Phys. A: Math. Gen., 38, (2005), R385-R406.}}

\vspace{1cm}

\end{center}

%%%%%%%%%%%%%%%%%%%%%%%%%%%%%%%%%%%%%%%%%%%%%%%%%%%%%%%%%%%%%%%%%%%%%%%%%%%%%%%%%%%%%%%%%%%
\begin{abstract}
The Hopf algebra structure underlying Feynman diagrams which
governs the process of renormalization in perturbative quantum
field theory is reviewed. Recent progress is briefly summarized
with an emphasis on further directions of research.
\\[0.5cm]
\end{abstract}

\smallskip

\newpage

%%%%%%%%%%%%%%%%%%%%%%%%%%%%%%%%%%%%%%%%%%%%%%%%%%%%%%%%%%%%%%%%%%%%%%%
\tableofcontents
%%%%%%%%%%%%%%%%%%%%%%%%%%%%%%%%%%%%%%%%%%%%%%%%%%%%%%%%%%%%%%%%%%%%%%%
%%%%%%%%%%%%%%%%%%%%%%%%%%%%%%%%%%%%%%%%%%%%%%%%%%%%%%%%%%%%%%%%%%%%%%%%%%%%%%%%%%%%%%%%%%%

\section{Introduction and Overview}

Quantum field theory (QFT) by now has a long and outstandingly
successful history in all theories of physics. Merging the two
major revolutionary achievements of early 20th century physics,
quantum mechanics and special relativity, the founding fathers of
QFT were setting out for an unified description of elementary
particles phenomena. Its ideas and techniques found far reaching
applications in different and very distinct areas of theoretical
physics, and pure and applied mathematics.

Several approaches to QFT have been developed so far. Wightman's
early axiomatic~\cite{StWh} setting leading to constructive QFT,
together with Haag's mathematically elegant and rigorous algebraic
formulation of QFT in terms of von Neumann algebras~\cite{Haag},
best describes the nowadays common believe of what should be the
general physical principles underlying any QFT. Still, despite the
enormous and mathematically rigorous progress which has been made
using these formulations, both approaches have several problems to
make fruitful contact with experimental results, whilst they give
a crucial insight into the structure of free quantum fields.

The perturbative approach to quantum field theory is the most
successful. Theoretical predictions of physical quantities made by
using their expansion in terms of --renormalized-- Feynman graphs
match experimental results with a vertiginous high precision.
Nevertheless, in most, if not all, of the interesting and relevant
4-dimensional quantum field theories, performing even simple
perturbative calculations one cannot avoid facing ill-defined
integrals. The removal of these divergences in a sound way is the
process of renormalization, better known by the illustrative
description of ''sweeping them under the carpet". The basic idea
of perturbative renormalization in QFT goes back to
Kramers~\cite{Brown}, and was successfully applied for the first
time in a 1947 seminal paper by Bethe~\cite{Bethe}, dealing with
the concrete problem of the self energy contribution for the Lamb
shift in perturbative quantum electrodynamics (QED). The latter
can nowadays be regarded as one of the best tested physics
theories. Its modern extension to the standard model of elementary
particles represents one of the cornerstones of our present
understanding of the physical world. Here again the perturbative
treatment together with renormalization is the bread-and-butter of
the practitioner in high energy physics.

Maintaining the physical principles of locality, unitarity, and
Lorentz invariance, renormalization theory may be summed up by the
statement that to all orders in perturbation theory the
(ultraviolet) divergencies can be absorbed in a redefinition of
the parameters defining the QFT. Here two distinct concepts enter,
that of renormalizability, and the process of renormalization. The
former distinguishes those theories with only a finite number of
parameters, lending them considerably more predictive power. The
process of renormalization instead works indifferently of the
number of parameters.

Soon after Bethe's paper on perturbative QED, there have been
several approaches to establish that quantum field theories are
renormalizable in general. Dyson~\cite{Dyson1,Dyson2} was the
first to do so, using integral equations and skeleton expansions
for Green's functions. His work was then continued by Salam and
Weinberg. Unfortunately, this attempt failed in the first
instance, due to a problem related to a particular 14th order QED
graph, but could be cured later. The second approach, based on
earlier work by St\"uckelberg and Green, was taken by Bogoliubov
and Parasiuk~\cite{BP,BS}, using a recursive subtraction method,
known as Bogoliubov's $\mathrm{\bar{R}}$-map. Also their proof
contained a loophole, but eventually found its final and
satisfying form with the work of Hepp~\cite{Hepp} and later
Zimmermann~\cite{Zimmermann}. This standard result is nowadays
well-known under the name Bogoliubov--Parasiuk--Hepp--Zimmermann
(BPHZ) renormalization prescription. Later, Epstein and
Glaser~\cite{EpGl} presented a rigorous proof of renormalizability
situated in the realm of the axiomatic treatment of QFT. A fourth
approach was taken by Blaer and Young~\cite{BY}, using the
renormalization group equations, going back to a suggestion by
Callan. At this point we refer the interested reader to consult
the work by Caswell and Kennedy~\cite{CaswellK},
Brown~\cite{Brown}, Delamotte~\cite{Delamotte},
Collins~\cite{Collins}, and Wightman~\cite{Wightman} for more
references and details.

Notwithstanding its somewhat notorious reputation, renormalization
theory, together with the gauge principle, forms the backbone of
the perturbative approach to physically relevant quantum field
theories. These days, the modern point of view, represented by the
concept of Wilson's renormalization group, elevates it even to a
fundamental structure in the understanding of high energy physics.

Unfortunately, despite its accomplishments, renormalization theory
was stigmatized, especially for its lack of a firm mathematical
underpinning. Indeed, examining the current introductory and
advanced literature on renormalization, as it is used in everyday
applications in many branches of physics, one feels the need for a
more conceptual picture unifying mathematical and computational
aspects. A possible reason for this situation might have been the
fact that its building blocks, the (one-particle irreducible)
Feynman graphs in itself appeared to be unrelated to a sound
mathematical structure that may underlie the renormalization
prescription in perturbative QFT.

Almost five decades after Bethe's work, this changed to a great
extend with the original paper by Kreimer~\cite{Kreimer1}
introducing the notion of Hopf algebra. The ensuing work by
Kreimer~\cite{Kreimer2,Kreimer6} and collaborators, especially
those of Broadhurst and Kreimer~\cite{BK1,BK2,BK3,BK4}, and Connes
and Kreimer~\cite{CK1,CK2,CK3} explored this new approach both in
terms of its mathematical and physical content, as well as its
computational aspects. The Hopf algebraic setting captures the
combinatorial and algebraic aspects of the process for
renormalization by organizing the Feynman graphs into a
combinatorial Hopf algebra, $\mathcal{H}_F$, which is a connected
graded commutative bialgebra, essentially characterized by its
non-cocommutative coproduct structure map. The formulation of
renormalization using Hopf algebras was completed in the work of
Connes and Kreimer. It gives rise to an elegant and useful
disentanglement of analytic and algebraic aspects of perturbative
renormalization in general QFT, affirming the remark that ''Few
physicists object nowadays to the idea that diagrams contain more
truth than the underlying formalism[...]" made by Veltman and 't
Hooft in \cite{tHV}.

In this review we will focus on an elementary introduction to the
Hopf algebra structure on Feynman graphs combined with the
description of a completely algebraic formulation of
renormalization in terms of a factorization problem valid for any
renormalization scheme, and based on a theorem by Atkinson and
Spitzer's identity~\cite{EGK1,EGK2}.

Let us continue with some more details. The restricted dual of the
Hopf algebra of Feynman graphs, denoted by $\mathcal{H}^*_{F}$,
contains the group $\mathcal{G}:=char(\mathcal{H}_F,\mathbb{C})$
of characters, that is, algebra homomorphisms from $\mathcal{H}_F$
to the underlying base field $\mathbb{C}$. Feynman rules are
understood as such linear and multiplicative maps, associating to
each Feynman graph, seen as a collection of vertices and edges,
its corresponding Feynman integral. This group of characters
possesses a corresponding Lie algebra of derivations, or
infinitesimal characters, $\mathcal{L}:=\partial
char(\mathcal{H}_F,\mathbb{C})$, which comes from a fundamental
pre-Lie algebra structure on Feynman graphs.

The ill-defined Feynman integrals are plagued with ultraviolet
divergences in general, and demand for a regularization
prescription, where we replace the base field $\mathbb{C}$ by a
(commutative and unital) algebra $A$ of Feynman amplitudes.
Alternatively, we might consider Taylor expansions on the level of
the integrands. Whichever way, this leads us to consider the space
of $A$-valued, or regularized, linear maps
$\Hom(\mathcal{H}_{F},A)$, which contains
$\mathcal{G}_A:=char(\mathcal{H}_{F},A)$, the group of regularized
characters, respectively its associated Lie algebra
$\mathcal{L}_A:=\partial char(\mathcal{H}_{F},A)$. As a principal
example serves dimensional regularization, where
$A:=\mathbb{C}[\varepsilon^{-1},\varepsilon]]$, the field of
Laurent series. In this context perturbative renormalization finds
a compact formulation as a factorization problem in the group
$\mathcal{G}_A$, to wit, the algebraic Birkhoff decomposition of
Feynman rules~\cite{CK1,CK2}. The initial proof of the
Connes--Kreimer factorization of regularized Feynman rules uses
the property that Laurent series actually form a commutative
Rota--Baxter algebra~\cite{Baxter,Rota1} with the pole part
projection, $R:=R_{ms}$, as linear Rota--Baxter operator (minimal
subtraction scheme map) fulfilling the Rota--Baxter relation (of
weight 1)
$$
  R(x)R(y)+R(xy)=R\big(R(x)y+xR(y)\big), \; \forall x,y \in A.
$$
The linearity of $R$ permits to define a unital, but now
non-commutative complete filtered Rota--Baxter algebra structure
on the space $\Hom(\mathcal{H}_{F},A)$, with convolution as
associative product. One of the fundamental results in the realm
of commutative Rota--Baxter algebras is Spitzer's classical
identity~\cite{EGK1,EGK2,RotaSmith,Spitzer}, and using its
generalization to non-commutative Rota--Baxter algebras, together
with Atkinson's factorization theorem~\cite{Atkinson} for
Rota--Baxter algebras, one can show that the multiplicative
factorization of Connes--Kreimer follows from an additive
decomposition through the exponential map~\cite{EGK1,EGK2}. Hereby
we realize Bogoliubov's $\mathrm{\bar{R}}$-map as a special case
of Spitzer's identity.\medskip

We hope that this brief review will guide the reader to crucial
aspects of the recent developments related to the Hopf algebraic
description of renormalization theory. The long list of references
is meant to indicate the rich spectrum of research directions
triggered by this approach. The modest mathematical style, i.e. we
do not strive for a rigorous theorem-proof presentation might help
the interested and novice reader to get a glimpse of the new
aspects which opened with the Hopf algebra point of view on
perturbative renormalization. Some of the remarks made during the
expository writing indicate points to be further developed. But,
we should underline that this article is neither meant to be an
introduction to (perturbative) quantum field theory nor to
renormalization theory in general. Rather, we would like to focus
on the by now well-understood and established
combinatorial-algebraic picture that makes renormalization theory
in perturbative QFT such a challenging and venerable subject from
both, mathematical and physical perspectives. After reading this
article and going back to the vast existing physics literature on
renormalization theory in all its facets, the reader may get an
idea of the interesting open questions related to its Hopf
algebraic description.

In the following we will comment on assorted references with
respect to their research directions in this field, in the hope to
facilitate access to this developing subject. Due to the review
character of this work and limited space none of these topics
could be treated in full detail. We start by mentioning two recent
papers \cite{FG3,Manchon} devoted in great detail to the general
Hopf algebra structure in renormalization theory. A more
mathematical, but shorter summary was given in~\cite{Boutet}.

The initial discovery of the Hopf algebra structure grew out of a
study of the number-theoretic properties of graphs with many
subdivergences, see \cite{Kreimer9}, which provides an overview of
some of the results up to the year 1999, including the link
between knot theory, Feynman graphs and number theory~\cite{BK5}.

The original work of Kreimer, and Connes and
Kreimer~\cite{CK1,CK2,CK3,CK4,Kreimer1,Kreimer2} explores and
settles the Hopf algebraic formulation of renormalization for
general perturbative QFT, and links it to non-commutative
geometry. It thereby establishes the Birkhoff decomposition for
Feynman rules giving rise to an unexpected correspondence with the
Riemann--Hilbert problem. In~\cite{CORV,FG1,FG2,GKM1,Kastler1,
KW,Kreimer4,RV} further details were given. Malyshev's
work~\cite{Malyshev1} shows the general character of Connes' and
Kreimer's combinatorial Hopf algebra, applying it to Riemann
surfaces in the context of ribbon Feynman graphs.

The link to Connes' non-commutative geometry becomes evident in
terms of a Hopf algebra of non-planar rooted trees~\cite{CK1},
solving a universal problem in Hochschild cohomology, and forming
the role model for the Hopf algebraic structure of
renormalization. This work renewed considerably the interest in
Hopf algebraic aspects of combinatorics such as rooted tree Hopf
algebras. In this context one must point out the work of
G.-C.~Rota~\cite{Rota4} and his school, especially Rota's and
A.~Joni's seminal work~\cite{JoniRota} from the late 1970ies,
forming the starting point for the theory of incidence Hopf
algebras, further developed in~\cite{Ehrenborg,Schmitt}, see also
\cite{FG3}. Holtkamp~\cite{Holtkamp}\footnote{See also L.~Foissy's
PhD-thesis.} showed that the non-commutative version of
Connes--Kreimer's Hopf algebra of rooted trees is isomorphic to
Loday and Ronco's~\cite{LodayRonco}. Aguiar and collaborators
explored in more detail rooted tree Hopf
algebras~\cite{AguBS,AguSot1,AguSot2}. Hoffman in~\cite{Hoffman}
improved a result of Panaite~\cite{Panaite}, showing the
isomorphism between the dual of Connes--Kreimer's Hopf algebra and
Grossmann--Larson's~\cite{G-L} rooted tree Hopf algebra (see
also~\cite{Foissy}). Brouder~\cite{Brouder} explored the relation
to Butcher's seminal work on Runge--Kutta integration
methods~\cite{Butcher}. Turaev in~\cite{Turaev1,Turaev2} extended
some combinatorial aspects of the Connes--Kreimer results,
especially with respect to the notion of pre-Lie coalgebras, and
thereby also gave a neat description of Connes--Kreimer's rooted
tree Hopf algebra.

Chapoton and Livernet~\cite{ChapotonLivernet} described free
pre-Lie algebras in terms of rooted tree operads. Mencattini and
Kreimer~\cite{KM1,KM2} further analyzed the insertion and
elimination Lie and pre-Lie algebraic structures of Feynman
graphs~\cite{CK4} in terms of infinite matrix representations.

In~\cite{EGK1,EGK2} the meaning of the Rota--Baxter relation in
the context of Connes--Kreimer's Birkhoff decomposition is
investigated in detail, pointing out some parallels to the theory
of classical integrable systems~\cite{BBT,STS1}. It thereby
provides the algebraic underpinning for the factorization in terms
of complete filtered Rota--Baxter algebras, Spitzer's identity and
Atkinson's multiplicative decomposition theorem for Rota--Baxter
algebras. This work was further extended in~\cite{EG2,EGGV}
describing the combinatorics of renormalization in terms of
unipotent triangular matrix representations, and their
factorization capturing the process of renormalization analogously
to the Birkhoff decomposition of Connes--Kreimer. An interesting
application of renormalization techniques and Rota--Baxter
algebras as described here to the iteration of symbols of
pseudodifferential operators can be found in~\cite{MP}.

The work of Broadhurst and Kreimer~\cite{BK1,BK2,BK3,BK4} develops
many computational and physics aspects. They show how to use the
coproduct structure of the Hopf algebra to efficiently compute the
forest formula and use the Hochschild cohomology of the algebra to
resum the perturbative series. The latter two of the
aforementioned references hence form the starting point for the
latest work of Kreimer~\cite{Kreimer7,KreimerDS}, and Bergbauer
and Kreimer~\cite{BergbauerKreimer2}, putting emphasis on
Hochschild cohomology of Hopf algebras as a source of locality,
the Dyson--Schwinger equations, and even the Slavnov--Taylor
identities for the couplings in generic gauge theories.

Finally, the authors in~\cite{GKM2,Malyshev2,Sakakibara} started
to analyze some aspects of renormalization group calculations in
the Hopf algebra context.

The work~\cite{BKK,BW} (see also~\cite{MUW}) is of more
computational character, indicating the efficiency of the use of
Hopf algebras in perturbative renormalization.

Several people~\cite{BergbauerKreimer1,Gudrun} investigated the
link between the Hopf algebra of renormalization to the most
rigorous approach to renormalization in perturbative QFT, provided
by the Epstein--Glaser prescription.

Recently, progress has been made in the mathematical context of
number theory, and motivic structures of Feynman integrals. The
notion of equisingular connections was used to explore Tannakian
categories and Galois symmetries in the spirit of differential
Galois theory in~\cite{CM1,CM2,CM3}. Underlying the notion of an
equisingular connection is the locality of counterterms, which
itself results from Hochschild cohomology. The resulting
Dyson--Schwinger equation allows for gradings similar to the
weight- and Hodge filtrations for the polylogarithm
\cite{Kreimer3,Kreimer8}. More concretely, the motivic nature of
primitive graphs has been established very recently by Bloch,
Esnault and Kreimer~\cite{BlochEK}.\medskip

Let us briefly outline the organization of the paper.
Section~\ref{section2} introduces briefly the basic Lie and Hopf
algebra structures in perturbative renormalization, including the
pre-Lie composition of Feynman graphs and Bogoliubov's
$\bar{\mathrm{R}}$-operation. The next section uses perturbative
QED as a simple example to manifest the aforementioned notions. In
Section~\ref{section4} we formulate the process of renormalization
in perturbative QFT as a factorization problem in purely algebraic
terms. Emphasis is put on the freedom in choosing a particular
regularization prescription, captured via the notion of
commutative unital Rota--Baxter algebra. Section~\ref{section5}
outlines the use of the Birkhoff decomposition introduced in the
former section on the level of diffeomorphisms of coupling
constants, in the realm of dimensional regularization together
with the minimal subtraction scheme as a particular useful
renormalization prescription. The review ends with a brief section
on the role of Hochschild cohomology in perturbative
renormalization. In an appendix we collect some general facts
about Rota--Baxter algebras as they form the main ingredient for
Section~\ref{section4}.

%%%%%%%%%%%%%%%%%%%%%%%%%%%%%%%%%%%%%%%%%%%%%%%%%%%%%%%%%%%%%%%%%%%%%%%%%%%%%%%%%%%%%%%%%%%

\section{From the Lie and Hopf algebras of graphs to Bogoliubov's formula}
\label{section2}

In this section we describe the elementary Lie and Hopf algebra
structures underlying perturbation theory. The reader looking for
a mathematical rigorous and detailed presentation of Hopf algebras
and related aspects is referred to the standard texts such
as~\cite{Abe,FGV,Kassel,Sw} (see also~\cite{FG3,Manchon}).

Let $\mathbb{K}$ be a field of characteristic zero. All
$\mathbb{K}$-algebras, denoted by a triple $(A,m,\eta)$, where $A$
is a $\mathbb{K}$-vector space with a product $m: A \otimes A \to
A$ and a unit map $\eta: \mathbb{K} \to A$, are supposed to be
associative. Similarly for coalgebras, denoted by the triple
$(C,\Delta, \bar{e})$, where the coproduct map $\Delta: C \to C
\otimes C$ fulfills coassociativity, i.e. $(\Delta \otimes \id)
\circ \Delta=(\id\otimes\Delta) \circ \Delta$, and $\bar{e}: C \to
\mathbb{K}$ is the counit map. The identity will be denoted by
$\One$. All algebra homomorphisms are supposed to be unital. A
bialgebra, denoted by a quintuple $\left( H =
\bigoplus_{i=0}^\infty H_{(i)}, m, \eta, \Delta, \bar{e} \right)$
consists of an algebra and coalgebra structure in a compatible
way. Here, $\Delta: H \to H \otimes H$ is the coproduct, $m: H
\otimes H \to H$ the product. These maps together with the counit
$\bar{e}: H\to \mathbb{K}$ fulfil the standard bialgebra axioms.
See the above general references for details. It is called
connected graded if $H_{(i)} H_{(j)} \subset H_{(i+j)}$ and
$\Delta(H_{(i)}) \subset \bigoplus_{j+k=i} H_{(j)} \otimes
H_{(k)},$ and if $\Delta(\One)=\One\otimes\One$ and
$H_{(0)}=\mathbb{K}\One$. The counit simply is $\bar{e}(\One) =
1\in\mathbb{K}$ and $\bar{e}=0$ on $\bigoplus_{i=1}^\infty
H_{(i)}.$

We call $\ker\bar{e}$ the augmentation ideal of $H$ and denote by
$P$ the projection $H \rightarrow \ker\bar{e}$ onto the
augmentation ideal, $P = \id - \eta\bar{e}.$ Furthermore, we use
Sweedler's notation $\Delta(h)=\sum h^\prime \otimes
h^{\prime\prime}$ for the coproduct. Let us define
 \begin{equation}
    {\rm Aug}^{(k)}= \big(\underbrace{P\otimes\cdots \otimes P}_{k\;{\rm times}}\big)\,
   \Delta^{k-1},\;\; H \to \{\ker\bar{e}\}^{\otimes k},
 \end{equation}
as a map into the $k$-fold tensor product of the augmentation
ideal. Here, $\Delta^{k-1}$ is defined inductively by
$\Delta^0:=\id$, and $\Delta^n:=(\Delta^{n-1}\otimes
\id)\circ\Delta$ for $n>0$. We let
\begin{equation}
    {\sl H}^{(k)}=\ker{\rm Aug}^{(k+1)}/\ker{\rm Aug}^{(k)},
\end{equation}
$\forall k \geq 1$. All bialgebras considered here are bigraded in
the sense that
\begin{equation}
    H=\bigoplus_{i=0}^\infty H_{(i)}=\bigoplus_{k=0}^\infty {\sl H}^{(k)},
\end{equation}
where $H_{(k)} \subset \oplus_{j=1}^k {\sl H}^{(j)}$ for all $k
\geq 1$. $H_{(0)}\simeq {\sl H}^{(0)}\simeq \mathbb{K}$.

While these algebraic notions may seem rather abstract, they
indeed govern the structure of quantum field theory. To understand
how, we have first to study the pre-Lie algebra structure of
one-particle irreducible Feynman graphs.

%%%%%%%%%%%%%%%%%%%%%%%%%%%%%%%%%%%%%%%%%%%%%%%%%%%%%%%%%%%%%%%%%%%%%%%%%%%%%%%
%%%%%%%%%%%%%%%%%%%%%%%%%%%%%%%%%%%%%%%%%%%%%%%%%%%%%%%%%%%%%%%%%%%%%%%%%%%%%%%

\subsection{The Pre-Lie Structure of Feynman Graphs}
\label{subsection1}

For each quantum field theory, we have an underlying free theory
which provides propagators and hence Feynman rules for edges in
Feynman graphs\footnote{We assume that the reader has already seen
a Feynman graph, otherwise he might take a brief look on Eq.
(\ref{pre-LieExam}).}. The request for local interactions and a
renormalizable theory then gives us Feynman rules for
interactions. With those graphs come the sets of one-particle
irreducible (1PI) graphs contributing to a chosen amplitude. The
amplitudes are distinguished by the external fields,
asymptotically free fields represented by external edges in the
graphs. We call this an external leg structure, denoted by
$\underline{r}$. For a renormalizable theory, there is a finite
number of such external leg structures, one for each monomial in
the Lagrangian.

For each such Feynman graph we hence have vertices as well as
internal and external edges. External edges are edges which have
an open end not connected to a vertex. They indicate the particles
participating in the scattering amplitude under consideration and
each such edge carries the quantum numbers of the corresponding
free field.  The internal edges and vertices form a graph in their
own right. For an internal edge, both ends of the edge are
connected to a vertex. For a graph $\Gamma$ we denote by
$\Gamma^{[0]}$  its set of vertices and by
$\Gamma^{[1]}:=\Gamma^{[1]}_{\rm int} \cup \Gamma^{[1]}_{\rm ext}$
its set of internal and external edges. Furthermore,
$\omega_{\underline{r}}$ is the number of space-time derivatives
appearing in the corresponding monomial in the Lagrangian.

We are considering 1PI Feynman graphs. By definition a graph
$\Gamma$ is 1PI if and only if all graphs, obtained by removal of
any one of its internal edges, are still connected. Such 1PI
graphs are naturally graded by their number of independent loops,
the rank of their first homology group
$H_{[1]}(\Gamma,\mathbb{Z})$. We write $|\Gamma|$ for this degree
of a graph $\Gamma$. Note that $|{\bf res}(\Gamma)|=0$, where we
let ${\bf res}(\Gamma)$, called the residue of $\Gamma$, be the
graph obtained when all edges in $\Gamma^{[1]}_{\rm int}$ shrink
to a point. The graph we obtain in this manner consists of a
single vertex, to which the edges $\Gamma^{[1]}_{\rm ext}$ are
attached. In case the initial graph was a self-energy graph, we
regard its residue as a single edge. We denote the set of all
external leg structures $\underline{r}$ by $\mathfrak{R}$. For a
renormalizable QFT it consists of the edges and vertices
corresponding to the monomials in the Lagrangian.

Having specified free quantum fields and local interaction terms
between them, one immediately obtains the set of 1PI graphs, and
can consider for a given external leg structure $\underline{r}$
the set $M_{\underline{r}}$ of graphs with that external leg
structure. The Green's function for the corresponding amplitude is
then obtained as the evaluation under the Feynman rules of the
formal sum
 \begin{equation}
    \Gamma^{\underline{r}}:= 1 + \sum_{{\bf res}(\Gamma)={\underline{r}}}
                                    \alpha^{|\Gamma|}\frac{\Gamma}{{\rm sym}(\Gamma)},
 \end{equation}
where we divide by the symmetry factor ${{\rm sym}(\Gamma)}$ and
$\alpha$ is a small parameter like (the square of) a coupling
constant. These sums exhibit rich structure thanks to the
algebraic structures of the single graphs \cite{KreimerDS} to be
discussed now.

For a renormalizable theory, we can define a superficial degree of
divergence
 \begin{equation}
    \omega= \sum_{\underline{r}\in \Gamma^{[1]}_{\rm int}\cup
            \Gamma^{[0]}}\omega_{\underline{r}}-4|H_{[1]}(\Gamma,\mathbb{Z})|,
 \end{equation}
for each such external leg structure: $\omega(\Gamma) =
\omega(\Gamma^\prime)$ if ${\bf res}(\Gamma) = {\bf
res}(\Gamma^\prime)$, all graphs with the same external leg
structure have the same superficial degree of divergence. Only for
a finite number of distinct external leg structures $\underline{r}
\in \mathfrak{R}$ will this degree indeed signify a divergence.
Our first observation is that there is a natural pre-Lie algebra
structure on 1PI graphs.

To see this, we define a bilinear operation on graphs
 \begin{equation}
    \Gamma_1 * \Gamma_2 = \sum_\Gamma n(\Gamma_1,\Gamma_2;\Gamma)\Gamma,
 \end{equation}
where the sum is over all 1PI graphs $\Gamma$. Here,
$n(\Gamma_1,\Gamma_2;\Gamma)$ is a section coefficient which
counts the number of ways a subgraph $\Gamma_2$ in $\Gamma$ can be
reduced to a point such that $\Gamma_1$ is obtained. The above sum
is evidently finite as long as $\Gamma_1$ and $\Gamma_2$ are
finite graphs, and the graphs which contribute necessarily fulfill
$|\Gamma| = |\Gamma_1| + |\Gamma_2|$ and ${\bf res}(\Gamma) = {\bf
res}(\Gamma_1)$.

One then has:
\begin{theorem}
The operation $\ast$ is pre-Lie:
    \begin{equation}
        [\Gamma_1\ast\Gamma_2]\ast \Gamma_3  -  \Gamma_1\ast[\Gamma_2\ast \Gamma_3] =
        [\Gamma_1\ast \Gamma_3]\ast \Gamma_2  - \Gamma_1\ast[\Gamma_3\ast \Gamma_2].
    \end{equation}
\end{theorem}
This is evident when one rewrites the $*$ product in suitable
gluing operations, using the dichotomy of inserting in nested or
disjoint manner. See~\cite{EMK,Kreimer5,Kreimer10,Kreimer11} for
more details.

Note that the equation claims that the lack of associativity in
the bilinear operation $\ast$ is invariant under permutation of
the elements indexed by $2,3$. This suffices to show that the
anti-symmetrization of this map fulfils the Jacobi identity. Hence
we get a Lie algebra ${\cal L}$ by anti-symmetrizing this
operation:
 \begin{equation}
    [\Gamma_1,\Gamma_2] = \Gamma_1\ast\Gamma_2-\Gamma_2\ast\Gamma_1.
    \label{Lie}
 \end{equation}
This Lie algebra is graded and of finite dimension in each degree.
Let us look at a couple of examples for pre-Lie products. We take
graphs from quantum electrodynamics (QED) as a rather self-evident
example. For the graphs $\gg$ and $\epemg$ with residues ${\bf
res}(\!\gg\!) = \!\!\begin{array}{c}\\[-.6cm] \scalebox{0.5}{\BOSONprop}\end{array}\!$
respectively ${\bf res}(\! \epemg
\!)=\!\!\begin{array}{c}\\[-.8cm] \!\scalebox{0.5}{\FERMprop}{}\end{array}\!$, we find
 \begin{eqnarray}
    \gg\ast \epemg & = & 2\ggv.
    \label{pre-LieExam}
 \end{eqnarray}
Together with ${\cal L}$ one is led to consider the dual of its
universal enveloping algebra ${\cal U}({\cal L})$ using the
theorem of Milnor and Moore~\cite{MilnorMoore}. For this we use
the above grading by the loop number.

This universal enveloping algebra ${\cal U}({\cal L})$ is build
from the tensor algebra
 \begin{equation}
    {\bf T}=\bigoplus_k T^k,\;T^k=\underbrace{{\cal L}\otimes\cdots\otimes {\cal L}}_{k\; {\rm times}},
 \end{equation}
by dividing out the ideal generated by the relations
 \begin{equation}
    a\otimes b - b \otimes a = [a,b] \in {\cal L}.
 \end{equation}
Note that in ${\cal U}({\cal L})$ we have a natural concatenation
product $m_*$. Even more, ${\cal U}({\cal L})$ carries a natural
Hopf algebra structure with this product. For that, the Lie
algebra ${\cal L}$ furnishes the primitive elements:
 \begin{equation}
    \Delta_*(a)=a\otimes 1+1\otimes a,\;\forall a\in {\cal L}.
 \end{equation}
It is by construction a connected finitely graded Hopf algebra
which is cocommutative but not commutative.

We can then consider its graded dual which will be a Hopf algebra
${\cal H}_{F}(m,\eta,\Delta,\bar{e})$ which is commutative but not
cocommutative. One finds the coproduct $\Delta$ upon using a
Kronecker pairing
 \begin{equation}
    <Z_\Gamma,\delta_{\Gamma^\prime}>= \begin{cases} 1,\;\Gamma=\Gamma^\prime \\
                                                        {0,\; {\rm
                                                        else}}.\end{cases}
 \end{equation}
From there, one determines all other structure maps with ease,
demanding that
 \begin{equation}
    \langle Z_{[\Gamma_2,\Gamma_1]},\delta_\Gamma \rangle =
        \langle Z_{\Gamma_1}\otimes Z_{\Gamma_2} - Z_{\Gamma_2}\otimes Z_{\Gamma_1},\Delta(\delta_\Gamma)\rangle.
 \end{equation}
In the above, we distinguished carefully between graphs $\Gamma$
as generators of the Lie algebra, denoted by $Z_\Gamma$, and
graphs $\Gamma$ as generators of the Hopf algebra, denoted by
$\delta_\Gamma$. The Lie algebra of graphs exponentiates to the
character group of the Hopf algebra as explained below, eventually
leading to Birkhoff factorization in that group.

The space of primitives of ${\cal U}({\cal L})$ is in one-to-one
correspondence with the set ${\rm Indec}({\cal H}_{F})$ of
indecomposables of ${\cal H}_{F}$, which is the linear span of its
generators.

${\cal H}_{F}$ is a connected graded commutative Hopf algebra
which describes renormalization theory. It operates on the
superficially divergent 1PI Feynman graphs of the theory. The
residues of these graphs are in one-to-one correspondence with the
terms in the Lagrangian of a given theory. Often it is the case
that several terms in a Lagrangian correspond to graphs with the
same number and type of external legs, but match to different
form-factor projections of the graph. In such cases, the above
approach can be easily adopted. Below in Section~\ref{section3} we
give an example for QED, incorporating its form-factor
decomposition into our approach.

%%%%%%%%%%%%%%%%%%%%%%%%%%%%%%%%%%%%%%%%%%%%%%%%%%%%%%%%%%%%%%%%%%%%%%%%%%%%%%%
%%%%%%%%%%%%%%%%%%%%%%%%%%%%%%%%%%%%%%%%%%%%%%%%%%%%%%%%%%%%%%%%%%%%%%%%%%%%%%%

\subsection{Bogoliubov's recursive subtraction formula}
\label{subsection2}

The above algebra structures are available once one has decided on
the set of 1PI graphs of interest. Those one-particle irreducible
graphs $\Gamma$ provide the generators $\delta_\Gamma$ of the Hopf
algebra ${\cal H}_{F}=\oplus_{i=0}^\infty \H_{(i)}$, where ${\cal
H}_{ F, {\rm{lin}}}:={\rm span}(\delta_\Gamma)$, with their
disjoint union providing the commutative product, which we denote
by juxtaposition.

Let $\Gamma$ be a 1PI graph. The  Hopf algebra ${\cal H}_{F}$
described above comes out to have a coproduct $\Delta:{\cal H}_{F}
\to {\cal H}_{F} \otimes {\cal H}_{F}$:
 \begin{equation}
    \Delta(\Gamma) = \Gamma \otimes \One  + \One \otimes\Gamma +
                            \sum_{\gamma{\subset}\Gamma}\gamma\otimes\Gamma/\gamma,
 \end{equation}
where the sum is over all unions of 1PI superficially divergent
proper subgraphs, and we extend this definition to products of
graphs,
$\Delta(\Gamma_1\Gamma_2)=\Delta(\Gamma_1)\Delta(\Gamma_2)$, so
that we get a bialgebra.

While the pre-Lie product respectively the Lie bracket inserted
graphs into each other, dually the coproduct disentangles them.
This is precisely what we make use of in renormalization theory:
we have to render each subgraph finite before we can construct a
local counterterm. Having a coproduct, two further structure maps
of ${\cal H}_{F}$ are immediate, the counit and the antipode. The
counit $\bar{e}$ vanishes on any non-trivial Hopf algebra element,
$\bar{e}(X)=0$, $X \neq \One$, but $\bar{e}(\One )=1$. The
antipode $S:{\cal H}_{F} \to {\cal H}_{F}$ is given by
 \begin{equation}
    S(\Gamma) = -\Gamma- \sum_{\gamma{\subset}
    \Gamma}S(\gamma)\Gamma/\gamma,
 \end{equation}
for $\Gamma \in \ker \bar{e}$, and $S(\One)=\One$. We can work out
examples for the coproduct of a graph:
 \begin{eqnarray}
    \Delta\big( \!\! \ggv \!\!\big) & = &\!\! \ggv\!\!\otimes \One + \One \otimes\!\! \ggv + 2\epemg\otimes \gg.
 \end{eqnarray}
And an antipode:
 \begin{equation}
    S\big(\!\!\!\! \begin{array}{l} \\[-0.4cm]
                          \epemfg \end{array} \!\!\!\!\big) = -\!\!\begin{array}{l} \\[-0.4cm]
                          \epemfg \end{array}  \!\!+ \gg\epem.
 \end{equation}
We note in passing that the gluing operation underlying the
pre-Lie insertion of graphs relies on gluing data which can be
reconstructed from the subgraphs $\gamma$ and cographs
$\Gamma/\gamma$ in the above coproduct. This is crucial in the
proof of locality of counterterms upon studying the Hochschild
cohomology of this Hopf algebra \cite{Kreimer10}.

We have by now obtained a Hopf algebra generated by combinatorial
elements, 1PI Feynman graphs. Its existence is indeed automatic
once one has chosen interactions and free fields.

As disjoint scattering processes give rise to independent
(divergent) amplitudes one is led to the study of regularized
characters of the Hopf algebra, to say $A$-valued maps $\phi:
{\cal H}_{F} \to A$ such that $\phi \circ m=m_A \circ (\phi
\otimes \phi)$. Here $A$ denotes a unital commutative algebra.

Usually, Hopf algebra characters, i.e., linear multiplicative
maps, assign to any element in the Hopf algebra an element in the
base field, and form a group under convolution, denoted by
$\mathcal{G}$. Motivated by the need for regularizing our theory,
due to ultraviolet (UV) divergencies showing up in higher loop
calculations, we take here a slightly more general point of view,
replacing the base field as target space, say $\mathbb{C}$, by a
suitable commutative and unital algebra, $A$, of -regularized-
Feynman amplitudes. The group of regularized, or $A$-valued, Hopf
algebra characters is denoted by $\mathcal{G}_A$, and the group
law is given by the convolution product
 \begin{equation}
    \phi_1 \star \phi_2 := m_A \circ (\phi_1 \otimes \phi_2) \circ \Delta,
    \label{def:convol}
 \end{equation}
so that the coproduct, counit and coinverse (the antipode) give
the product, unit and inverse of this group, as befits a Hopf
algebra.

The study of tree-level amplitudes in lowest order perturbation
theory justifies to assign to each edge a propagator and to each
elementary scattering process a vertex which define the Feynman
rules $\phi({\rm \bf res}(\Gamma))$ and the underlying Lagrangian,
on the level of residues of these very graphs. With the Feynman
rules providing a canonical character $\phi$, we will have to make
one further choice: a renormalization scheme. The need for such a
choice is no surprise: after all we are eliminating short-distance
singularities in the graphs which renders their remaining finite
part ambiguous, albeit in a most interesting manner.

We choose a $\mathbb{K}$-linear map  $R:A \to A$, from which we
obviously demand that it does not modify the UV-singular
structure, i.e., $R^2=R$, and furthermore that it obeys
 \begin{equation}
    R(x)R(y) + R(xy) = R\big(R(x)y) + R(xR(y)\big),
    \label{RB}
 \end{equation}
an equation which guarantees the multiplicativity of
renormalization and lies at the heart of the Birkhoff
decomposition which emerges below: it tells us that elements in
$A$ split into two parallel subalgebras given by the image and
kernel of $R$. Algebras for which such a map exists fall into the
class of Rota--Baxter algebras, well-known in mathematics. See
below in Section \ref{section4} (and Appendix
\ref{appendix:RotaBaxter}) for more details .

Let us take a shortcut for the moment and see how all the above
structure comes together in renormalization theory. Starting with
a regularized Feynman rules character $\phi$, we define a further
character $S_ R^\phi$ which deforms $\phi \circ S$, the inverse of
the $\phi$, slightly and delivers the counterterm for $\Gamma$ in
the renormalization scheme $R$:
 \begin{equation}
 \label{counterterm}
    S_R^\phi(\Gamma)=-R\big[m_A(S_R^\phi\otimes\phi\circ P)\Delta(\Gamma)\big]=
                     -R[\phi(\Gamma)]-R\left[\sum_{\gamma{\subset}\Gamma}
                     S_R^\phi(\gamma)\phi(\Gamma/\gamma)\right],
 \end{equation}
for $\Gamma$ in $\ker\bar{e}$. Comparing with the undeformed
inverse of $\phi$
 \begin{equation}
  \phi \circ S(\Gamma) = m_A(\phi \circ S \otimes\phi \circ P)\Delta(\Gamma)=
                         -\phi(\Gamma)-\sum_{\gamma{\subset} \Gamma}\phi\circ S (\gamma)\phi(\Gamma/\gamma)
 \end{equation}
allows to easily understand finiteness of renormalized quantities,
thanks to the independence of counterterms on kinematical
variables. Later, in Section~\ref{section4} we will fully derive
Equation (\ref{counterterm}) and the results below from a more
mathematical point of view from the fact that $R$ is a
Rota--Baxter operator.

We conclude that $S_R^\phi$ is an element of the group of
regularized characters, ${\cal G}_{A}$, of the Hopf algebra,
$S_R^\phi\in {\rm Spec}({\cal G}_{A})$. We now have determined the
renormalized Lagrangian:
 \begin{equation}
    Z^{\underline{r}}=S_R^\phi(\Gamma^{\underline{r}}).
 \end{equation}
The standard results of renormalization theory follow immediately
using the group of regularized characters: the renormalization of
a graph $\Gamma$ is obtained by the application of a renormalized
character, $S_R^\phi \star \phi$
 \begin{equation}
    S_R^\phi\star\phi(\Gamma)=m_A(S_R^\phi\otimes\phi)\Delta(\Gamma)
 \end{equation}
for $\Gamma \in \ker \bar{e}$, and Bogoliubov's
$\bar{\mathrm{R}}$-operation is obtained as
 \begin{equation}
 \label{BogosBARmap}
   \bar{\mathrm{R}}(\Gamma)=m_A(S_R^\phi\otimes\phi)({\rm id}\otimes P)\Delta(\Gamma)=
                                                      \phi(\Gamma)+ \sum_{\gamma{\subset}\Gamma}
                                                            S_R^\phi(\gamma)\phi(\Gamma/\gamma).
 \end{equation}
In the following we write $\bar{R}(\Gamma)=:\bar{\phi}(\Gamma)$,
so that we have
 \begin{equation}
    S_R^\phi\star\phi(\Gamma)=\bar{\phi}(\Gamma) + S_R^\phi(\Gamma).
 \end{equation}
$S_R^\phi\star\phi$ is an element in the group of regularized
characters, $\mathcal{G}_A$, of the Hopf algebra. This Lie group
has indeed the previous Lie algebra ${\cal L}$ of graph insertions
as its Lie algebra: ${\cal L}$ exponentiates to ${\cal G}_A$.

What we have achieved at this moment is a local renormalization of
quantum field theory. Let $m_{\underline{r}}$ be a monomial in the
Lagrangian $L$ of degree $\omega_{\underline{r}}$,
 \begin{equation}
    m_{\underline{r}}=D_{\underline{r}}\{\varphi\},
 \end{equation}
for some suitable derivation $D_{\underline{r}}$ on the fields
$\varphi$. Then one can prove using the Hochschild cohomology and
induction over the augmentation degree in ${\cal H}_F$:
\begin{theorem} (locality)
 \begin{equation}
    Z^{\underline{r}}D_{\underline{r}}\{\varphi\}=D_{\underline{r}}Z^{\underline{r}}\{\varphi\},
 \end{equation}
renormalization commutes with infinitesimal space-time variations
of the fields.
\end{theorem}

Let us finally give the renormalization of a Feynman graph, say
$\Gamma=\!\!\ggv$. \allowdisplaybreaks{
 \begin{eqnarray}
     \Delta\big(\!\!\ggv\!\!\big) & = & \!\!\ggv\!\otimes \One+\One\otimes\!\ggv+2\epemg\otimes\gg \nonumber\\
 \bar{\phi}\big(\!\!\ggv\!\!\big) & = & \phi\big(\!\!\ggv\!\!\big)+
                                            2S^\phi_R\big(\!\epemg\!\big)\phi\big(\!\gg\!\big) \nonumber\\
                          & = & \phi\big(\!\!\ggv\!\!\big)-
                                            2R\left[\phi\big(\!\!\epemg\!\!\big)\right]\phi\big(\!\gg\!\big) \nonumber\\
   S_R^\phi\big(\!\!\ggv\!\!\big) & = & -R\left[\bar{\phi}\big(\!\!\ggv\!\!\big)\right] \label{renorm}\\
     \phi_R\big(\!\!\ggv\!\!\big) &\equiv & S_R^\phi\star\phi\big(\!\!\ggv\!\!\big)=
                                   [{\rm id}-R]\circ\left[\bar{\phi}\big(\!\!\ggv\!\!\big)\right].\nonumber
 \end{eqnarray}}

%%%%%%%%%%%%%%%%%%%%%%%%%%%%%%%%%%%%%%%%%%%%%%%%%%%%%%%%%%%%%%%%%%%%%%%%%%%%%%%
%%%%%%%%%%%%%%%%%%%%%%%%%%%%%%%%%%%%%%%%%%%%%%%%%%%%%%%%%%%%%%%%%%%%%%%%%%%%%%%

\section{Example: QED}
\label{section3}

The QED Lagrangian (density) for an electron coupled to the
electromagnetic field in coordinate space  reads (we skip the
$1/2(\partial \cdot A)^2$ term)
\begin{equation}
  \label{def:QED}
  L_{QED}(\alpha,m) = i\bar{\psi} \partial\!\!\!/ \psi
            + \bar{\psi} eA\!\!\!/ \psi
            + m \bar{\psi} \psi
            + \frac{1}{4} F^2,
\end{equation}
with the electromagnetic field tensor $F_{\mu\nu}:=\partial_\mu
A_\nu - \partial_\nu A_\mu$ and $F^2:=F_{\mu\nu}F^{\mu\nu}$. The
Euler-Lagrange equations for this term give the Maxwell equations.
For the first and third term they give the Dirac equation. We use
units defined by $c=\hbar=1$, for which the elementary charge
squared, $e^2=4\pi \alpha_{QED}$. It is a dimensionless quantity
depending on the fine-structure constant $\alpha:=\alpha_{QED}
\simeq \frac{1}{137}$. The first term in (\ref{def:QED}) describes
the kinetic energy of the electron, and together with the mass
term, $m \bar{\psi} \psi$, it constitutes the free Lagrangian
density for an electron. The second term in (\ref{def:QED})
describes the minimal coupling of the electron with the
electromagnetic field. The first two terms are dictated by local
gauge invariance of the QED Lagrangian, i.e., invariance with
respect to multiplying the electron field $\psi$ by a position
dependent phase factor.  Let us introduce for every term in
$L_{QED}$ a 1PI Green's function $G^{r}(p^2,\alpha,m,\mu)$, with
\begin{equation}
 \label{def:QEDmonomials}
  r \in \left\{   i\bar{\psi} \partial\!\!\!/\psi ,\
                         \bar{\psi}e A\!\!\!/ \psi,\
                                m \bar{\psi} \psi ,\
                                 \frac{1}{4}F^2
                                \right\},
\end{equation}
all of which transform as scalars under the Lorentz group.

Usually, 1PI Green's functions of QED are given in standard
notation by the expressions
\begin{enumerate}
    \item $\Gamma_{\nu}(p_1,p_2,m,\alpha,\mu)$ the vertex function;

    \item $S_{\rm F}^{-1}(p,m,\alpha,\mu)$ the inverse fermion propagator;

    \item $P^{-1}_{\nu\tau}(p,m,\alpha,\mu)$ the inverse photon
      propagator,
\end{enumerate}
\noindent all depending on the bare parameters mass $m$ and
coupling constant $\alpha$, and 't Hooft's parameter $\mu$ which
sets the scale for the one-parameter groups of automorphisms of
the Lie algebra which run the renormalization group \cite{CK2}. An
extra parameter, such as 't Hooft's unit mass $\mu$, enters
naturally in the context of regularization, and is needed for
dimensional reasons~\cite{Collins}. From symmetry considerations
we find the following form factor decompositions, where we now
assume the vertex at zero momentum transfer for simplicity:
\begin{enumerate}
\label{def:formfactors}
 \item $
 \Gamma_{\nu}(p,p,m,\alpha,\mu)
       = e G^{\bar{\psi} A\!\!\!/ \psi}(p^2,m,\alpha,\mu)\gamma_{\nu}
         + eH^{\bar{\psi} A\!\!\!/ \psi}(p^2,m,\alpha,\mu)\frac{p\!\!/ p_{\nu}}{p^2}
$;

\item
$
 S^{-1}(p,m,\alpha,\mu)=G^{\bar{\psi} p\!\!/ \psi}(p^2,m,\alpha,\mu)p\!\!/
                        + G^{ m\bar{\psi} \psi}(p^2,m,\alpha,\mu)m\mathbb{I}
$;

\item
$
 P^{-1}_{\nu\tau}(p,m,\alpha,\mu)
       = \Pi_{\rm tr}^{\frac{1}{4} F^2}(p^2,m,\alpha,\mu)\left\{g_{\nu\tau}p^2 - p_\nu p_\tau \right\}
          + \Pi_{\rm long}^{\frac{1}{2}(\partial\cdot A)^2}(p^2,m,\alpha,\mu)p_\nu p_\tau
$ ,
\end{enumerate}
\noindent reintroducing a longitudinal term for completeness. Let
us introduce a graphical notation for the monomials of the QED
Lagrangian (\ref{def:QED}), which will form the building blocks of
our graphical Hopf algebra of Feynman graphs.
\begin{equation*}
 \bar{\psi} \partial\!\!\!/ \psi \
   \longleftrightarrow
 \begin{array}{ccc}
    \FERMprop\\[-0.1cm]
    {\tiny{\partial}}
 \end{array}
 \qquad\quad
   m \bar{\psi} \psi \
   \longleftrightarrow
 \begin{array}{c}
  \FERMprop\\[-0.3cm]
   {\tiny{m}}
 \end{array}
 \qquad\quad
  \begin{array}{cc}
             &\\
       \bar{\psi} eA\!\!\!/ \psi \ \longleftrightarrow & \\ [-0.9cm]
                                                       & \QEDvertex
  \end{array}
\end{equation*}
The first two graphs on the left represent the electron propagator
parts, corresponding to the derivation and mass contribution in
(\ref{def:QED}), respectively. The third graph is the QED vertex,
representing the interaction of the electromagnetic field with
fermions in (\ref{def:QED}). To ease the notation we suppressed
spinorial indices.

Next we have the transversal respectively longitudinal parts of
the bosonic photon propagator,
$$
%%%%%%%%%%%%%BOSONprop-transversal
  \frac{1}{4} F^2\
   \longleftrightarrow
 \begin{array}{c}
  \BOSONprop\\[-0.3cm]
   {\tiny{ t}}
 \end{array}
\qquad \quad
%%%%%%%%%%%%%BOSONprop-longitudional
 1/2(\partial \cdot A)^2\
 \longleftrightarrow
 \begin{array}{c}
  \BOSONprop\\[-0.3cm]
   {\tiny{\rm long}}
 \end{array}
$$
In the following, we will work with a transversal photon
propagator for concreteness. Once we have the graphical notation
we define the -coordinate space- QED Feynman rules
$\widehat{\phi}$ such that for any $\underline{r} \in
\mathfrak{R}_{QED}$
\begin{equation}
  \label{def:QEDgraphs}
  \mathfrak{R}_{QED}:=\left\{
\begin{array}{cccc}
  \begin{array}{ccc}
    \FERMprop\\[-0.1cm]
    {\tiny{\partial}}
  \end{array},
  &
  \begin{array}{c}
    \FERMprop\\[-0.3cm]
    {\tiny{m}}
  \end{array},
  &
  &,\
 \begin{array}{c}
           \\[-0.2cm]
  \BOSONprop\\[-0.3cm]
   {\tiny{t}}
%   \BOSONprop\\[-0.2cm]
%   {\tiny{l}}
 \end{array} \\[-1.3cm]
 & & \QEDvertex\!\! &
\end{array}\right\}
\end{equation}
we get back the corresponding coordinate space QED Lagrange
monomial $r$ in (\ref{def:QEDmonomials})
\begin{equation}
  \widehat{\phi}(\underline{r}) = r.
\end{equation}
We write the QED Lagrangian (\ref{def:QED}) pictorially
\begin{equation}
\label{def:graphQED}
  L_{QED}(\alpha,m) = \sum_{\underline{r}\in \mathfrak{R}_{QED} }
           \widehat{\phi}(\underline{r}).
\end{equation}
With the QED free propagators and vertex in $\mathfrak{R}_{QED}$
at hand we have available the one-particle irreducible Feynman
diagrams which provide amplitudes corresponding to these
propagations and interaction, as before. We can now introduce
partitions of unity for the form-factor decomposition of any
Green's function we are interested in. For example, if we do want
to decompose the self-energy of the fermion into its kinetic
energy and mass part \allowdisplaybreaks{
\begin{eqnarray}
        G^{\bar{\psi} p\!\!/ \psi}(p^2,m,\alpha,\mu) & = &
                                \frac{1}{p^2} \mathrm{Tr}\big(p\!\!/ S^{-1}(p,m,\alpha,\mu)\big) \\
        G^{ m\bar{\psi} \psi}(p^2,m,\alpha,\mu) & = &
                                \frac{1}{m}\mathrm{Tr}\big(S^{-1}(p,m,\alpha,\mu)\big).
\end{eqnarray}}
We can easily incorporate this by promoting our Hopf algebra to
pairs $(\Gamma,\sigma)$~\cite{CK1}, where $\sigma$ indicates the
desired form-factor obtained by composing the Feynman rules with a
suitable projector as above.

The sum over all projectors defines a partition of unity
 \begin{equation}
        {\rm id}=\sum_\sigma P_\sigma.
 \end{equation}
This structure can be easily incorporated on the level of Hopf
algebras, generalizing the study of external structures by setting
for the pairs $(\Gamma,\sigma)$
 \begin{equation}
        \Delta(\Gamma,\sigma)=\sum (\Gamma^\prime,1)\otimes (\Gamma^{\prime\prime},\sigma).
 \end{equation}
Note that should we wish we can partition the unity on the left
hand side above, \begin{equation}(\Gamma^\prime,1)\to
(\Gamma^\prime,\sigma_R)\end{equation} if we want to use
information that only particular form-factors $\sigma_R$ need
renormalization. Under the Feynman rules these pair of graphs then
evaluate to the amplitudes corresponding to the structure
functions defined by the projectors signified by the indicated
external leg structures. The resulting Hopf algebras for such
pairs $(\Gamma,\sigma)$ are decorated versions of the ones for
graphs only, and define graph-like structures very similar to the
tree-like structures of Turaev for the Hopf algebra of rooted
trees~\cite{Turaev1}. Examples can be found in
\cite{Delbourgo-Kreimer,BK1}.

As an example, we might wish to renormalize the mass part of
$\epemfg$, using the knowledge that the photon self-energy is
transversal. We hence work out the coproduct
 \begin{equation}
            \Delta\big((\!\!\!\!\begin{array}{l} \\[-0.4cm]
                          \epemfg \end{array}\!\!\!\!,\sigma_m)\big)= (\!\!\!\!\begin{array}{l} \\[-0.4cm]
                          \epemfg \end{array}\!\!\!\!,\sigma_m)\otimes \One+
           \One\otimes(\!\!\!\!\begin{array}{l} \\[-0.4cm]
                          \epemfg \end{array}\!\!\!\!,\sigma_m)+(\gg,\sigma_{\rm trans})\otimes (\epem,\sigma_m).
 \end{equation}
Under the Feynman rules, we evaluate using the corresponding
projectors and obtain the expected Lorentz scalar structure
functions and counterterms following the routine as outlined in
Eqs. (\ref{renorm}).

%%%%%%%%%%%%%%%%%%%%%%%%%%%%%%%%%%%%%%%%%%%%%%%%%%%%%%%%%%%%%%%%%%%%%%%%%%%%%%%
%%%%%%%%%%%%%%%%%%%%%%%%%%%%%%%%%%%%%%%%%%%%%%%%%%%%%%%%%%%%%%%%%%%%%%%%%%%%%%%

\section{Renormalization as a factorization problem}
\label{section4}

As we have seen the notion of connected graded commutative Hopf
algebra appears naturally in the context of perturbative
renormalization of 1PI Feynman graphs. Both, composing Feynman
graphs in terms of the pre-Lie insertion product, where we replace
vertices by Feynman graphs with compatible external leg structure,
as well as their decomposition by eliminating subgraphs, i.e.,
replacing non-trivial 1PI subgraphs by their residues, look very
familiar when inspecting the subtraction procedure encoded in the
original BPHZ prescription~\cite{BP,BS,CaswellK,Collins}. The
later was invented to extract the finite part of the Feynman
integral corresponding to a Feynman graph via a regularized
Feynman rules character, while maintaining fundamental physical
principles, such as locality, unitarity, and Lorentz invariance.

The commutative Hopf algebra of Feynman graphs, $\mathcal{H}_{F}$,
and its graded dual,
$\mathcal{H}^*_{F}=\mathrm{Hom}(\mathcal{H}_{F},\mathbb{C})$ are
intimately related by the Milnor--Moore theorem. The space
$\mathcal{H}^*_{F}$ together with the convolution product and the
counit map $\bar{e}: \mathcal{H}_{F} \to \mathbb{C}$ as unit forms
a unital, associative and non-commutative $\mathbb{C}$-algebra,
which contains the group of characters,
$\mathcal{G}:=char(\mathcal{H}_{F},\mathbb{C})$, i.e., linear
functionals $\phi \in \mathcal{H}^*_{F}$ from $\mathcal{H}_{F}$ to
$\mathbb{C}$ respecting multiplication, $\phi(\Gamma_1
\Gamma_2)=\phi(\Gamma_1)\phi(\Gamma_2)$, $\Gamma_1$, $\Gamma_2 \in
\mathcal{H}_{F}$. This group of multiplicative maps possesses a
corresponding Lie algebra, $\mathcal{L}=\partial
char(\mathcal{H}_{F},\mathbb{C}) \subset \mathcal{H}^*_{F}$, of
derivations, or infinitesimal characters, i.e., linear maps $Z \in
\mathcal{H}^*_{F}$, satisfying Leibniz' rule
$$
  Z(\Gamma_1 \Gamma_2)= Z(\Gamma_1)\bar{e}(\Gamma_2)
                                  +\bar{e}(\Gamma_1) Z(\Gamma_2)
$$
for all $\Gamma_1$, $\Gamma_2 \in \mathcal{H}_{F}$. The grading of
$\mathcal{H}_{F}$ implies a decreasing filtration on
$\mathcal{H}^*_{F}$, which allows us to introduce a metric, and
therefore a distance map. $\mathcal{H}^*_{F}$ is complete with
respect to the induced topology. The exponential map
$\mathrm{exp}^{\star}$ gives a bijection between the Lie algebra
$\mathcal{L}$ and its corresponding group $\mathcal{G}$.

Using QED as an example we have seen that in general Feynman rules
for any perturbative QFT form a subclass of characters. Also, we
had to face the severe problem that the associated Feynman
integrals for graphs beyond the tree level suffer from ultraviolet
divergencies in the limit of large momenta, or equivalently small
distances. Therefore, one is forced to invoke a regularization of
such integrals, or more generally the Feynman rules themselves.
Actually, there is no specific selection rule for such a
regularization, indeed one must assure that the final physical
result is independent of such an unphysical intermediate step. At
the same time it is of vital importance that the regularization
prescription used in calculations respects as many physical
properties of the underlying theory as possible, such as gauge
symmetries. We will ignore such subtleties and take the following
stance. In the above Hopf algebraic setting, the regularization of
our theory is achieved by replacing the base field $\mathbb{C}$ as
target space of maps in $\mathcal{H}^*_{F}$ by an unital algebra
$A$, of which we demand commutativity, and the existence of a
linear map $R$ satisfying the Rota--Baxter relation
\begin{equation}
 \label{eq:RBrel}
 R(x)R(y)+R(xy)=R\big(R(x)y\big)+\big(xR(y)\big)
\end{equation}
for all $x,y \in A$.  For $R$ being such a Rota--Baxter map,
$\tilde{R}:=\id_A - R$ also satisfies relation~(\ref{eq:RBrel}).
Such algebras are well-known in mathematics under the name
Rota--Baxter algebra (see Appendix~\ref{appendix:RotaBaxter} for
more details). As a principal example we mention here dimensional
regularization, where the image of Feynman rules lives in the
field of Laurent series,
$A=\mathbb{C}[\varepsilon^{-1},\varepsilon]]$, with the pole part
projection $R:=R_{ms}$ as Rota--Baxter map. In the examples one
encounters in QED renormalization calculations, where the
regularization could have been for instance a simple cut-off of
the momentum integrals, the map $R$ is given in terms of an
evaluation at a specific fixed momentum (on-shell scheme),
$R_{q}(f)(p):=f(q)$, and trivially satisfies relation
(\ref{eq:RBrel}), as it is an idempotent algebra homomorphism,
i.e., the zeroth order in the Taylor expansion of the map $f$ at
point $q$.

The space $\A:=\mathrm{Hom}(\mathcal{H}_{F},(A,R))$ of $A$-valued
linear functionals, together with the convolution product
(\ref{def:convol}) and unit $e:=u_A \circ \bar{e}$, $e(\One)=1_A$,
forms an associative non-commutative algebra, containing the group
of regularized characters, $\mathcal{G}_A :=
char(\mathcal{H}_{F},A)$, and its corresponding Lie algebra,
$\mathcal{L}_A := \partial char(\mathcal{H}_{F},A)$, of
regularized derivations. The linearity of the Rota--Baxter map $R$
on the regularization target space $A$, gives rise to a
Rota--Baxter algebra structure on
$\mathrm{Hom}(\mathcal{H}_{F},(A,R))$, induced in terms of the
linear map $\R$, which is defined for any $f \in \A$, by $\R(f):=R
\circ f \in \A$. As before, we can equip $\A$ with a decreasing
filtration of Rota--Baxter ideals
$$
  \A=\A_0 \supset \A_1 \supset \dots \supset \A_n \supset \dots
$$
making it a complete filtered non-commutative Rota--Baxter algebra
with convolution product as composition, $(\A,\R,\{\A_n\}_{n\geq
1})$, since $\R(\A_n) \subset \A_n$ for all $n$. Here we have
$\mathcal{L}_A$ as a Lie subalgebra of $\A_1$, and $\mathcal{G}_A$
is a subgroup of $\widehat{\mathcal{G}}:=e +\A_1$, such that
\begin{eqnarray}
 \exp^{\star}: & \A_1 \to e+\A_1,\quad \exp^{\star}(Z):=
                                    \sum_{n=0}^\infty\frac{Z^{\star n}}{n!},     \label{eq:exp} \\
 \log^{\star}: & e+\A_1 \to \A_1,\quad \log^{\star}(e+Z):=
                                   -\sum_{n=1}^\infty \frac{(-Z)^{\star n}}{n} \label{eq:log}
\end{eqnarray}
are well-defined with respect to convolution and inverse to each
other. Furthermore $\exp^{\star}$ restricts to a bijection between
$\mathcal{L}_A$ and $\mathcal{G}_A$.

Atkinson's~\cite{Atkinson} (see
Appendix~\ref{appendix:RotaBaxter}) factorization theorem for
associative Rota--Baxter algebras implies in the above setting,
that for a fixed $\phi = e + Z \in \mathcal{G}_A$ the solutions $X
\in e+\mathcal{R}(\A_1)$, $Y \in e+\tilde{\mathcal{R}}(\A_1)$ of
the equations
\begin{equation}
 \label{eq:At1}
 X = e - \R(X \star Z) \ \;\mathrm{ resp.}\ \; Y = e - \tilde{\R}(Z \star Y)
\end{equation}
solve the factorization problem
\begin{equation}
 \label{eq:At2}
 e + Z = \phi = X^{-1} \star Y^{-1},
\end{equation}
which can be easily checked. If the  Rota--Baxter map $R$ is
idempotent, the decomposition in (\ref{eq:At2}) is unique. In the
following we denote $\phi_{-}:=X$ and $\phi_{+}=:Y^{-1}$.
Spitzer's classical identity~\cite{Spitzer} for commutative
Rota--Baxter algebras can be generalized to non-commutative
Rota--Baxter algebras, thereby implying one of the main results of
the Hopf algebraic approach to renormalization in QFT~\cite{CK1},
to wit the algebraic Birkhoff decomposition of Connes and Kreimer,
which we formulate as a theorem.

\begin{theorem}{\rm{\cite{EGK2}}}
Let $\H_{F}$ be a connected graded Hopf algebra of Feynman graphs
associated with a perturbatively treated renormalizable QFT. Let
$A$ be a commutative unital Rota--Baxter algebra with an
idempotent Rota--Baxter operator $R$. Let $\A$ be the complete
filtered algebra $\Hom(\H_{F},A)$. \allowdisplaybreaks{
\begin{enumerate}
\item $(\A,\R,\{\A_n\}_{n\geq 1})$ is a complete filtered
Rota--Baxter algebra with the idempotent operator $\R(f):=R \circ
f$. \label{it:cRBA}
\item For $\phi = e + Z \in \mathcal{G}_A$, there is unique
$\phi_- \in e +\R(\A_1)$ and $\phi_+ \in e +\tilde{\R}(\A_1)$ such
that
\begin{equation}
  \label{Birkhoff}
  \phi = \phi_-^{-1} \star \phi_+.
\end{equation}
\label{it:decom}
\item The elements $\phi_-=e-\R(\phi_{-}\star(\phi-e))$ and
$\phi_+=e-\tilde{\R}( \phi_{+} \star (\phi^{-1} - e))$ solving Eq.
(\ref{Birkhoff}) take the following form for $\Gamma \in \ker
\bar{e}$:
\begin{align}
 \phi_-(\Gamma)&= -R\big(\phi(\Gamma)+\sum_{(\Gamma)}\phi_-(\Gamma')\phi(\Gamma'')\big),\label{special1}\\
 \phi_+(\Gamma)&=
 \tilde{R}\big(\phi(\Gamma)+\sum_{(\Gamma)}\phi_-(\Gamma')\phi(\Gamma'')\big). \label{special2}
%\\
%              &= \tilde{R}(\phi(\Gamma)+\sum_{(\Gamma)}\phi(\Gamma')\phi_+(\Gamma''))
\end{align}
\label{it:rec}
\item Spitzer's identity for non-commutative complete Rota--Baxter
algebras implies that the linear maps $\phi_-$ and $\phi_+$ can be
written as
\begin{equation}
 \label{eq:Spi}
  \phi_- = \exp^{\star}\Big( -\R\big(\chi(Z) \big)\Big) \ \mathrm{ resp.} \quad
  \phi_+ = \exp^{\star}\Big( \tilde{\R}\big(\chi(Z)\big)\Big)
\end{equation}
and naturally give algebra homomorphisms. \label{it:decgp}
\end{enumerate}}
\label{thm:algBi}
\end{theorem}
Equation (\ref{special2}) follows by general arguments for
Rota--Baxter algebras, as we can write for $\phi_{+}$ the
following equation
$$
 \phi_{+}=e + \tilde{\R}(\phi_{+} \star (e-\phi^{-1}))
         =e + \tilde{\R}(\phi_{-} \star (\phi-e)).
$$
The unique map $\chi: \A_1 \to \A_1$ in (\ref{eq:Spi}) is the
key-result for the generalization of Spitzer's identity to
non-commutative Rota--Baxter algebras, and satisfies the equation
\begin{equation}
 \chi(Z)= Z - BCH\Big(\R\big(\chi(Z)\big),\tilde{\R}\big(\chi(Z)\big)\Big),
 \label{BCH-recur}
\end{equation}
where $BCH(x,y)$ denotes the Baker--Campbell--Hausdorff relation
$$
 \exp(x)\exp(y)=\exp\big(x+y+BCH(x,y)\big).
$$
The non-linear map $\chi$ was introduced in~\cite{EGK1,EGK2}, and
was called the $BCH$-recursion. The reader may find it helpful to
consult~\cite{EG2} for more details, and~\cite{Manchon} for a more
conceptual proof in the context of Lie algebras.

As a proposition to this theorem, we mention without giving
further details the fact that Bogoliubov's
$\bar{\mathrm{R}}$-operation (\ref{BogosBARmap}) can be written as
an exponential using the double Rota--Baxter convolution product,
$\star_{\R}$, on $(\A,\R,\{\A_n\}_{n\geq 1})$ (see Eq.
(\ref{def:doubleProd}) in Appendix \ref{appendix:RotaBaxter}):
$$
  \bar{\mathrm{R}}(\Gamma)=\bar{\phi}(\Gamma)= \phi_{-}\star
  (\phi-e)(\Gamma)=
  -\exp^{\star_{\R}}\big(-\chi(Z) \big)(\Gamma)
$$
for $\Gamma \in \ker \bar{e}$. Finally let us mention that the
above notion of complete Rota--Baxter algebra and Theorem
\ref{thm:algBi} becomes very transparent for uni- and nilpotent
upper (or lower) triangular matrices with entries in a commutative
Rota--Baxter algebra~\cite{EG2,EGGV}.

The above theorem presents a purely algebraic setting for the
formulation of renormalization as a factorization problem in the
group of regularized Hopf algebra characters, situated in the
theory of non-commutative Rota--Baxter algebras with idempotent
Rota--Baxter map. The formulae for the counterterm
(\ref{special1}) and renormalized character (\ref{special2}) are
completely dictated by a general decomposition structure, which
characterizes Rota--Baxter algebras~\cite{Atkinson}. The
additional property of $R$ being a projector implies a direct
decomposition of the algebra, hence the uniqueness of the
factorization in (\ref{Birkhoff}). We would like to emphasis the
necessary freedom in the choice of the regularization
prescription, encoded in the particular structure of the
commutative Rota--Baxter algebra $A$ as target space of linear
Hopf algebra functionals in $\Hom(\H_{F},A)$.

Specializing the target space Rota--Baxter algebra $A$ in the
above theorem to the field of Laurent series, i.e., using
dimensional regularization, we recover the original setting
in~\cite{CK1}, opening a hitherto hidden geometric viewpoint on
perturbative renormalization in terms of a correspondence to the
Riemann--Hilbert problem. This approach was further extended
in~\cite{CM1,CM2,CM3}.

%%%%%%%%%%%%%%%%%%%%%%%%%%%%%%%%%%%%%%%%%%%%%%%%%%%%%%%%%%%%%%%%%%%%%%%
%%%%%%%%%%%%%%%%%%%%%%%%%%%%%%%%%%%%%%%%%%%%%%%%%%%%%%%%%%%%%%%%%%%%%%%

\section{Diffeomorphisms of physical parameters}
\label{section5}

In the above, we obtained a unique Birkhoff decomposition of
Feynman rules $\phi \in Spec({\cal G}_A)$ into two characters
$\phi_- =: S_R^\phi \in Spec({\cal G}_A)$ and $\phi_+ = S_R^\phi
\star \phi \in Spec({\cal G}_A)$, for any idempotent Rota--Baxter
map $R$. Thanks to Atkinson's theorem this is possible for any
renormalization scheme $R$. For the minimal subtraction scheme it
amounts to the decomposition of the Laurent series
$\phi(\Gamma)(\varepsilon)$, which has poles of finite order in
the regulator $\varepsilon$, into a part holomorphic at the origin
and a part holomorphic at complex infinity. This has a geometric
interpretation upon considering the Birkhoff decomposition of a
loop around the origin, providing the clutching data for the two
half-spheres defined by that very loop, which is central in the
work of Connes and Kreimer \cite{CK2,CK3}. The geometric
interpretation leads to motivic Galois theory upon studying the
equisingularity of the corresponding connection in the
Riemann--Hilbert correspondence \cite{CM1,CM2}, itself a result of
the Hochschild cohomology of these Hopf algebras
\cite{BergbauerKreimer1,BergbauerKreimer2}.

Our understanding of each term in the perturbative expansion and
its renormalization have found hence satisfying mathematical
interpretations. The character group ${\cal G}_A$ is a poorly
understood object though, it is far too big. Fortunately
renormalization can be captured by the study of diffeomorphisms of
physical parameters, as by the very definition the range of
allowed modifications in renormalization theory is the variation
of the coefficients of monomials $\hat{\phi}(\underline{r})$ of
the underlying Lagrangian
 \begin{equation}
    L=\sum_{\underline{r}\in \mathfrak{R}}Z^{\underline{r}}\, \hat{\phi}(\underline{r}).
 \end{equation}
We can now eliminate the use of ${\cal G}_A$ as one can regain the
Birkhoff decomposition at the level of diffeomorphisms of the
coupling constants.

One proceeds by using that renormalized couplings provide a formal
diffeomorphism
 \begin{equation}
    g_{\rm new}=g_{\rm old}\;Z^g,
 \end{equation}
where
 \begin{equation}
    Z^g= \frac{Z^v}{\prod_{e \in {\bf res}(v)^{[1]}_{\rm ext}}\sqrt{Z^e}},
 \end{equation}
for some vertex $v$, which obtains the new coupling in terms of a
diffeomorphism of the old. This formula provides indeed a Hopf
algebra homomorphism from the Hopf algebra of diffeomorphisms to
the Hopf algebra of Feynman graphs, regarding $Z^g$, a series over
counterterms for all 1PI graphs with the external leg structure
corresponding to the coupling $g$, in two different ways: it is at
the same time a formal diffeomorphism in the coupling constant
$g_{\rm old}$ and a formal series in Feynman graphs. As a
consequence, there are two competing coproducts acting on $Z^g$.
Their consistency defines the required homomorphism, which
transposes to a homomorphism from the largely unknown group of
regularized characters of ${\cal H}_F$ to the one-dimensional
diffeomorphisms of this coupling. Hence one concludes \cite{CK3}:

\begin{theorem} Let the unrenormalized effective coupling constant $g_{\rm eff}
(\varepsilon)$ viewed as a formal power series in $g$ be
considered as a loop of formal diffeomorphisms and let $g_{\rm
eff} (\varepsilon)= (g_{{\rm eff}_-})^{-1}(\varepsilon) \, g_{{\rm
eff}_+}(\varepsilon)$ be its Birkhoff decomposition in the group
of formal diffeomorphisms. Then the loop $g_{{\rm eff}_-}
(\varepsilon)$ is the bare coupling constant and $g_{{\rm eff}_+}
(0)$ is the renormalized effective coupling.
\end{theorem}

%%%%%%%%%%%%%%%%%%%%%%%%%%%%%%%%%%%%%%%%%%%%%%%%%%%%%%%%%%%%%%%%%%%%%%%
%%%%%%%%%%%%%%%%%%%%%%%%%%%%%%%%%%%%%%%%%%%%%%%%%%%%%%%%%%%%%%%%%%%%%%%

\section{The role of Hochschild cohomology}
\label{section6}

The Hochschild cohomology of the Hopf algebras of 1PI graphs
illuminates the structure of 1PI Green's functions in various
ways:
 \begin{itemize}
    \item   it gives a coherent proof of locality of counterterms --the very fact that
     \begin{equation}
             [Z^{\underline{r}},D_{\underline{r}}]=0,
     \end{equation}            the coefficients in the Lagrangian remain independent of momenta,
            and hence the Lagrangian a polynomial expression in fields and
            their derivatives; \cite{BergbauerKreimer2,Kreimer10,KreimerDS}

    \item   the quantum equation of motions take a very succinct form
            identifying the Dyson kernels with the primitives of the Hopf
            algebra \cite{BergbauerKreimer2,Kreimer10,KreimerDS},
            and hence replacing a sum over all graphs by a sum over all primitive graphs;

    \item   sub-Hopf algebras emerge from the study of the Hochschild cohomology
            which connect the representation theory of these Hopf algebras to the
            structure of theories with internal symmetries leading to the
            Slavnov--Taylor identities for the couplings~\cite{KreimerDS};

    \item   these Hopf algebras are intimately connected to the structure of
            transcendental functions like the generalized polylogarithms which
            play a prominent role these days ranging from applied particle
            physics to recent developments in mathematics, in particular the primitive
            graphs which provide the Dyson kernels allow for a motivic interpretation~\cite{BlochEK}.
 \end{itemize}
For more information, we refer the reader to the literature
indicated.

\appendix

\section{Basic facts about general Rota--Baxter algebras}
\label{appendix:RotaBaxter}

For the reader's convenience we collect some basic notions of more
mathematical nature concerning mainly Rota--Baxter operators, in
the hope that from the above presentation they become redundant.
For more details we refer the reader to the standard literature,
e.g.~\cite{Atkinson,Guo1,RotaSmith}. Rota--Baxter operators (also
known as Baxter operators in older mathematical references) were
an active field of mathematical research in the late 1960s and
early 1970s. After an almost three decades long period of dormancy
they reappeared, as if on cue, in the mathematical literature in
the context of dendriform algebras~\cite{Aguiar,K1,JLL}, number
theory~\cite{Guo2}, generalizations of shuffle
products~\cite{EG1,GuoKeigher}, and Hopf
algebras~\cite{AndrewsGuo}, as well as in theoretical physics in
the seminal work of Kreimer and collaborators on the Hopf algebra
of renormalization~\cite{CK1,EGK1,EGK2,Kreimer2}.
\smallskip

In the following $\mathbb{K}$ denotes the base field of
characteristics zero, over which all algebraic structures are
defined. In general an algebra always means an associative unital
$\KK$-algebra, not necessarily commutative. The algebra unit is
simply denoted by $1$.\medskip

In Section \ref{section4} we encountered Rota--Baxter operators
respectively the Rota--Baxter relation in the context of
renormalization schemes, i.e. subtraction operators for the BPHZ
method. Let $A$ be an algebra together with a linear endomorphism
$R: A \to A$. We call the tuple $(A,R)$ a Rota--Baxter algebra of
weight $\theta \in \KK$, if the map $R$ fulfills the Rota--Baxter
relation (of weight $\theta$)
\begin{equation}
  \label{eq:RBrelation}
  R(x)R(y)+\theta R(xy)=R\big(R(x)y + xR(y)\big)
\end{equation}
for all $x,y \in A$. Without proof we state the fact that the
operator $\tilde{R}:=\theta \id_A - R$ is a Rota--Baxter operator
of weight $\theta$, too, such that the mixed relation
\begin{equation}
  \label{eq:mixedRBrel}
  R(x) \tilde{R}(y)=\tilde{R}\big(R(x)y\big) + R\big(x\tilde{R}(y)\big)
\end{equation}
is satisfied for all $x,y \in A$. The map $B:=\theta\id_A-2R$
satisfies the modified Rota--Baxter relation
\begin{equation}
  B(x)B(y)+\theta^2 xy=B\big(B(x)y + xB(y)\big).
\end{equation}
For $\theta \neq 0$, the normalized map $\theta^{-1} R$ is a
Rota--Baxter operator of weight one. Therefore without lost of
generality we may suppose in the following the canonical weight
one case. A Rota--Baxter (left-) right-ideal $I$ is a (left-)
right-ideal $I$ of $A$ such that $R(I) \subseteq I$. A
Rota--Baxter ideal is a Rota--Baxter left- and right-ideal.

The American mathematician Glen Baxter introduced this relation
1960 in his probability studies in fluctuation
theory~\cite{Baxter}. Later, the Italian born American
mathematician Gian-Carlo Rota~\cite{RotaSmith,Rota1,Rota2,Rota3},
and others~\cite{Cartier,Kingman}, notably F.V.
Atkinson~\cite{Atkinson}, explored in detail Baxter's work from
different perspectives in analysis, algebra and combinatorics. The
case $\theta = 0$ corresponds to the integration by parts property
of the usual Riemann integral $I: \F\to\F$, $I[f](x) := \int_0^x
f(t)\,dt$ in the algebra $\F$ of continues functions on $\RR$, to
wit,
\begin{equation}
    I[f_1]\,I[f_2] = I\bigl[ I[f_1] f_2 + f_1 I[f_2] \bigr],
\label{eq:ach-so}
\end{equation}
for $f_1,f_2 \in \F$. We already encountered the pole part
projection $R_{ms}$ in dimensional regularization as an example of
an idempotent Rota--Baxter map of weight one. The images of $R$ as
well as $\tilde{R}$ form subalgebras in $A$. Let $R$ be a
projector on $A$. For $R$ to satisfy the Rota--Baxter relation is
equivalent to a direct decomposition of $A=R(A)\oplus
\tilde{R}(A)$. This is just the special case of Atkinson's
additive decomposition theorem~\cite{Atkinson}, characterizing a
general Rota--Baxter algebra $(A,R)$ as a subdirect difference of
the images of $R$ and $\tilde{R}$.

The Lie algebra associated to $(A,R)$, with standard commutator
bracket forms a Rota--Baxter Lie algebra, $(\L_A,R)$, with $R$
fulfilling
\begin{equation}
   \label{eq:RBLieRelation}
   [R(x),R(y)]+\theta R([x,y])=R\big([R(x),y] + [x,R(y)]\big),
\end{equation}
better known as (operator) classical Yang--Baxter equation. Let us
mention here that it was rediscovered in this form in the early
1980ies by some Russian physicists in the context of classical
integrable systems (see e.g.~\cite{BBT,STS1} for references and
more details). This curious coincidence of Baxter and Baxter just
happens to reveal the connections of Rota--Baxter operators with
many areas of mathematics and physics.

The vector space underlying $A$, equipped with the new product
\begin{equation}
  \label{def:doubleProd}
  x \star^{(1)}_R y:=xR(y) + R(x)y - xy
\end{equation}
is again a Rota--Baxter algebra with Rota--Baxter map $R$, which
we denote $(A_1,R)$. Hence, all Rota--Baxter algebras $(A,R)$,
associative or non-associative, come with a whole hierarchy of
so-called double Rota--Baxter algebras, $(A_n,R)$, $n\in\NN$.
Relation (\ref{eq:RBrelation}) naturally implies that $R(x
\star^{(1)}_R y)=R(x)R(y)$, i.e. $R$ is an algebra homomorphism
from $A_1$ to $A$, or more generally from $A_n$ to $A_{n-1}$.

Baxter's original motivation for his work was to prove that for a
commutative algebra $A$ together with a linear map $R$ satisfying
relation (\ref{eq:RBrelation}), that later bore his name, the
following identity for fixed $a \in A$ \allowdisplaybreaks{
\begin{eqnarray}
  \exp\Big(- \frac{1}{\theta}R\big(\log(1 - \theta at)\big) \Big)
         &=&\sum_{n=0}^\infty t^n \underbrace{R\big(R(R( \cdots
         R}_{n-times}(a)a\dots)a)a\big)
%         \underbrace{R\big(aR(a(\dots aR(a)\dots))\big)}_{dd}
  \label{eq:SpitzerId}
\end{eqnarray}}
holds in the formal power series ring $\A:=A[[t]]$, with $t$ being
a commuting parameter. This famous relation is called Spitzer's
classical identity, and appeared in Frank Spitzer's 1956
paper~\cite{Spitzer}. In~\cite{EGK1} this was generalized to
non-commutative Rota--Baxter algebras, based on a BCH-type
recursion formula, the key result we used earlier, see Eq.
(\ref{BCH-recur}). A more general setting in which the above
factorization may be regarded, and of which the algebra $A[[t]]$
is just a special case, is that of complete filtered Rota--Baxter
algebras~\cite{EGK2,EG2,EGGV}.

The right hand side of (\ref{eq:SpitzerId}) is the unique solution
of the recursive equation
\begin{equation}
   \label{eq:Atkinson1}
   X= 1 + tR(Xa).
\end{equation}
This is a natural generalization of the recursion $f=1+I[gf]$
corresponding to the differential equation $f'=gf$, $f(0)=1$,
solved by $\exp(I[g])$, in $(\F,I)$ where $I$ is the Riemann
integral. It was again Atkinson~\cite{Atkinson} in 1963, who
observed that for any Rota--Baxter algebra, not necessarily
commutative, a solution to Eq. (\ref{eq:Atkinson1}), and its
companion equation for $\tilde{R}$
\begin{equation}
   \label{eq:Atkinson2}
   Y= 1 + t\tilde{R}(aY)
\end{equation}
in $A[[t]]$, solve the multiplicative decomposition problem
\begin{equation}
   \label{eq:AtkinsonFact}
   (1-\theta a) = X^{-1}Y^{-1},
\end{equation}
for any element $a \in A$. $R$ being a projector implies a unique
decomposition of the element $(1-\theta a) \in A$. This genuine
factorization property of Rota--Baxter algebras is further
analyzed in~\cite{EG2}.

%%%%%%%%%%%%%%%%%%%%%%%%%%%%%%%%%%%%%%%%%%%%%%%%%%%%%%%%%%%%%%%%%%%%%%%
%%%%%%%%%%%%%%%%%%%%%%%%%%%%%%%%%%%%%%%%%%%%%%%%%%%%%%%%%%%%%%%%%%%%%%%

\vspace{0.5cm} {\emph{Acknowledgements}}: The first author is very
thankful for a PhD grant provided by the Ev.~Studienwerk Villigst,
as well as for an extra travel grant which allowed him to visit
BIRS (Canada). Also, he would like to thank warmly the Theory
Department at the Physics Institute of Bonn University for
unlimited support and encouragement. Both authors would like to
thank the Fields Institute, where this work was finished, for warm
hospitality. The collaboration with Prof. L.~Guo is acknowledged.

%%%%%%%%%%%%%%%%%%%%%%%%%%%%%%%%%%%%%%%%%%%%%%%%%%%%%%%%%%%%%%%%%%%%%%%
%%%%%%%%%%%%%%%%%%%%%%%%%%%%%%%%%%%%%%%%%%%%%%%%%%%%%%%%%%%%%%%%%%%%%%%

%%%%%%%%%%%%%%%%%%%%%%%%%%%%%%%%%%%%%%%%%%%%%%%%%%%%%%%%%%%%%%%%%%%%%%%%%%%%%%%%%%%%%%%%%

\end{document}